\documentclass[aps,superscriptaddress,showpacs,nofootinbib,eqsecnum]{revtex4} 
                   
\usepackage{amssymb}  
\usepackage{epsf} 
\usepackage{eucal} 
\usepackage{epsfig}
\usepackage{graphicx}

\begin{document}

\title{Dissipation Coefficients from Scalar and Fermion Quantum
Field Interactions}

\author{Mar Bastero-Gil} \email{mbg@ugr.es} \affiliation{Departamento de
  F\'{\i}sica Te\'orica y del Cosmos, Universidad de Granada, Granada-18071,
  Spain}

\author{Arjun Berera} \email{ab@ph.ed.ac.uk} \affiliation{SUPA, 
School of Physics
and Astronomy, University of Edinburgh, Edinburgh, EH9 3JZ, United Kingdom}

\author{Rudnei O. Ramos} \email{rudnei@uerj.br} \affiliation{Departamento de
  F\'{\i}sica Te\'orica, Universidade do Estado do Rio de Janeiro, 20550-013
  Rio de Janeiro, RJ, Brazil} 

\newcommand{\be}{\begin{equation}} \newcommand{\ee}{\end{equation}}
\newcommand{\bea}{\begin{eqnarray}} \newcommand{\eea}{\end{eqnarray}}

\begin{abstract}
 
Dissipation coefficients are calculated in the
adiabatic, near thermal equilibrium regime for a large class of
renormalizable interaction configurations involving a two-stage
mechanism, where a background scalar field is coupled to heavy
intermediate scalar or fermion fields which in turn are coupled to
light scalar or fermion radiation fields. These interactions are 
typical of warm inflation microscopic model building.
Two perturbative regimes are shown where well defined approximations
for the spectral functions
apply. One regime is at high temperature, when the masses of both intermediate
and radiation fields are less than the temperature scale and where the
poles of the spectral functions dominate. The other regime is at low
temperature, when the intermediate field masses are much bigger than
the temperature and where the low energy and low three-momentum regime
dominate the spectral functions. 
The dissipation coefficients in these
two regimes are derived. However, due to resummation issues for the high 
temperature case, only phenomenological approximate 
estimates are provided for the dissipation in this regime. 
In the low temperature case, higher loop contributions are suppressed
and so no resummation is necessary.
In addition to
inflationary cosmology, the application of our results
to cosmological phase transitions
is also discussed.

\bigskip \bigskip

{\bf In Press JCAP (2011).}

\end{abstract}

\pacs{98.80.Cq, 11.10.Wx}

\maketitle

\section{Introduction}
                  
Many physical processes involving adiabatic dissipative 
effects near thermal equilibrium occur in relativistic systems of
bosonic and fermionic quantum fields.  In the evolution of the early
universe, several phase transition are believed to have occurred, with
Grand Unified, Electroweak, quantum cromodynamics (QCD),  
and chiral amongst the most commonly
studied, and for both first and second order type transitions.  Many
physical implications are associated with these transitions including
baryogenesis, defect formation, magnetic field formation, and relic
particle production.  In all such cases, interactions of the order
parameter, typically a scalar field, with other fields will lead to
various dissipative effects, and these effects, though not
always accounted for, can have significant influence on physical
processes.

An example of a phase transition in which dissipation has significant
influence  and even changes the basic dynamical nature of the process
has been seen in the inflationary universe scenario.  This scenario
involves the evolution of a scalar field which during the process of
changing phases of the system also induces inflationary expansion of
the universe.  In the standard picture of inflation \cite{ci},
the inflaton field  is modelled such that the interactions with other
fields  play no role during inflationary expansion.  This leads to a
thermodynamic supercooled phase in the universe during inflation.
After this, to end inflation and put the  universe back into a
radiation dominated phase, the inflaton enters a period of coherent
oscillations, where interactions with other fields now become
important to reheat the universe with particles \cite{reheatu}.  An
alternative to this cold inflation picture is to consider dynamics
where the inflaton interacts with other fields all during inflation.
In such a case, dissipative effects become important and the dynamics
of inflation is completely altered.  In this warm inflationary
dynamics \cite{Berera:1995ie}, radiation production occurs concurrent
with inflationary expansion, and the seeds of density perturbation are
thermal in origin \cite{im,Berera:1995wh,Berera:1999ws}, thus
significantly altering the observable signatures of inflation compared
to cold inflation.

Our primary motivation for the dissipative coefficients calculated
in this paper is for
their application to warm inflationary cosmology.  However,
our results may also have
a broader range of applications in different contexts involving the
nonequilibrium dynamics of fields, and we will briefly
discuss these possibilities later in the paper.

Typical warm inflation models involve a
two-stage field interaction mechanism, where a scalar field is coupled to
another intermediate  field which in turn is coupled to yet another
decay field.  The intermediate field (which will be called the
``catalyst'' field) is a heavy field (in this paper we  consider it
either a scalar or a fermion) that can decay  into light fields (which
will be called ``radiation'' fields and will here again be either
scalars and/or fermions).  These types of interactions, although
naturally emerge in the  hierarchy of interactions, were first
realized for calculations of dissipation  coefficients
from model building applications to warm inflation and the interaction
structure was generically termed as the two-stage mechanism \cite{BR1}
(for recent reviews, please see \cite{BMR,BasteroGil:2009ec}).  
These two-stage
interactions, although more complicated than the direct interactions,
where the relevant field modes would directly decay in light radiation 
particles, offer some simplifying features. For one thing, in the regime where
the intermediate field mass is heavy relative  to the temperature
whereas the radiation fields are light, what we term the
low-temperature regime, we can show that leading order contributions
suffice in general to give a reliable estimate for the dissipation
coefficient. This may not be the case in the regime where both the 
intermediate and final
field masses are small relative to the temperature, what we term the
high-temperature regime,  where a different coupling governs the vertex
compared to the decay width and the appearance of infrared and near 
on-shell divergences preclude the need for resummation of higher order terms,
as we will detail later on.

In this paper we compute dissipation
coefficients coming from intermediate fields, either scalars or
fermions, in two-stage field interacting models, in the adiabatic,
near thermal equilibrium regime.  
Within this work, we develop a very new
approximation which we call the low-momentum approximation. 
We show in this paper, both analytically and
numerically that this low-momentum approximation is valid for the
low-temperature regime, where a heavy mass field is involved in
mitigating the interaction between the background field and the light
fields. This low-momentum approximation
was first applied in \cite{BMR,mossxiong0}. In this paper we will 
further extend those earlier results and give a detailed analysis of the
possible different temperature regimes that can be defined in terms
of the different energy scales involved.   

To contrast the analysis in this paper of this new low-momentum
approximation, there is another very common approximation
used in the literature, which picks out the poles
of the propagators as dominating the integral expressions of 
dissipative coefficients. We will refer to this as the pole
approximation.  One key result we will show in this
paper is for these sorts of two-stage processes where the
scalar background field interacts with the light radiation fields via
heavy catalyst fields, in the low-temperature regime, defined
as where the heavy catalyst field masses are much bigger than the
temperature scale, this new low-momentum approximation dominates
over the pole approximation. In particular, if one extended the
pole approximation down into this low-temperature regime, one
would find exponentially damped behavior for the dissipative coefficients.
However, since the low-momentum approximation dominates this
effect, the result is that the dissipative coefficients
damp much slower, as a power law with respect to temperature,
and so are much bigger in the regime of low-temperature, as compared to
the results that would be obtained from the pole approximation.

This paper
focuses only on renormalizable two-stage interactions  between scalar
and fermion fields.   A thorough analysis of dissipation
coefficients for many of the applications stated above, also require
treating gauge fields, which is beyond the scope of this paper, but we
plan to examine them in a future paper.  Nevertheless, even in gauge
theories, there are interactions of the sorts computed in this paper,
which up to now have not been treated in the literature.

The paper is organized as follows. In Sect. \ref{diss} we briefly
review the physics and definitions leading to the appearance of dissipative
terms in the effective equations of motion for background fields.  
In Sect. \ref{interactions} we specify all the
interaction Lagrangian densities which we will treat in the paper.
Particular attention is given to the interactions relevant to the
two-stage mechanism \cite{BR1},  where dissipative effects are
determined by the decay of a heavy intermediate field  into light
fields, with the former coupled to a relevant background field, whose
dynamics we are interested in.  The formal expressions for calculation
of  the dissipative coefficients, within the model
interactions and decay forms we considered in our derivations, are
given in Sect. \ref{dissvis2stage}. In Sect. \ref{partprod} we give a
physical interpretation of the dissipation coefficients by showing
their  connection with particle production.  In Sect. \ref{specfunc}
the spectral functions, which are the fundamental quantities needed
for calculating the dissipation coefficients, are
computed. The results for dissipation  coefficients in
the very low, low  and high-temperature regimes are presented in
Sect. \ref{results}.  Various applications of these coefficients to
inflation and phase transitions in general are examined in
Sect. \ref{applications}. Section \ref{summary} gives our summary and
concluding remarks. Two Appendices are also  included to give
technical details. In Appendix \ref{models} we show possible model
building realizations  of the type of interactions we consider in this
paper.  In Appendix \ref{app} we give some of the details for the
calculations of the  imaginary parts of the relevant self-energy
contributions considered in this paper  and the corresponding
particles decay widths.

\section{Dissipation Coefficients}
\label{diss}

In this section we describe how dissipation  coefficients are
defined in quantum field theory. We start by discussing the definition and
derivation of the dissipation coefficient in a generic interacting quantum
field theory model.  

A system, displaced from a state of equilibrium, when interacting with  an
environment,  is expected to exhibit non-unitary evolution.  The system
dynamics will typically experience dissipative effects as it returns to its
equilibrium state.    A common example involving quantum fields  is the case
of some background scalar field $\varphi$ that is interacting with other
fields, with $\varphi$ having an amplitude that is initially displaced from
equilibrium. Let us describe this field theory model as having standard
kinetic terms and with a potential given by

\begin{equation}
V(\varphi, X) = V(\varphi) + V_{\rm int}(\varphi, X)\;,
\label{VphiX}
\end{equation}

\noindent
where $X$ represent any other field or degree of freedom that is coupled to
the background field and $V_{\rm int}(\varphi, X)$ gives how this background
field is coupled to this other field.  A proper study of the evolution of
this background field can be performed in the context of the in-in, or the
closed-time path (CTP) functional formalism (for an introduction, see for
example \cite{bellac}). By integrating over the $X$ field, a nonlocal
effective equation of motion for $\varphi$ can be derived.  If we write the
interaction term in Eq. (\ref{VphiX}) as

\begin{equation}
V_{\rm int}(\varphi, X) = f(\varphi) g(X)\;,
\end{equation}

\noindent
then an ensemble averaged effective equation of motion for the system,
described by the background field  $\varphi$, has the generic form up to first
order in the nontrivial nonlocal  terms (see also Ref.~\cite{calzetta+hu} for
a recent review)

\begin{eqnarray}
\partial_\mu \frac{\partial {\cal L}_{\rm eff,r}[\varphi]} {\partial
  (\partial_\mu \varphi)} - \frac{\partial   {\cal L}_{\rm eff,r}[\varphi]}
        {\partial \varphi} - i \frac{\partial f(\varphi)}{\partial \varphi}
        \int d^4 x' \theta(t-t') \left[ f(\varphi(x')) - f(\varphi(x)) \right]
        \langle [ g(X(x)),g(X(x')) ] \rangle =  0\;,
\label{effeom}
\end{eqnarray}

\noindent
where ${\cal L}_{\rm eff,r}[\varphi]$ is the renormalized effective Lagrangian
density for $\varphi$ and $\langle \cdots \rangle$ are ensemble averages with
respect to an equilibrium (quantum or thermal) state.  These ensemble averages
over the fields can also be conveniently expressed in terms of the causal
two-point Green's functions for the fields $X$, as shown e.g. in
Refs. \cite{GR,BGR,BR1,BMR,BR21,BR22}, with the form of the nonlocal term in
Eq. (\ref{effeom}) dependent on the form for the interaction term $V_{\rm
  int}(\varphi, X)$.  A simple case is for example when we have a bi-quadratic
interaction of the scalar field $\varphi$, assumed homogeneous, $\varphi
\equiv \varphi(t)$,  with some other scalar field represented by $X$, 

\begin{equation}
V_{\rm int}(\varphi, X) = \frac{g_1^2}{2} \varphi^2 X^2\;.
\end{equation}

\noindent
In this case, the nonlocal term in Eq. (\ref{effeom}) becomes~\cite{GR}

\begin{eqnarray}
&&- i \frac{\partial f(\varphi)}{\partial \varphi} \int d^4 x' \theta(t-t')
  \left[ f(\varphi(x')) - f(\varphi(x)) \right] \langle [ g(X(x)),g(X(x')) ]
  \rangle  \nonumber \\ && = 2 g_1^4 \varphi(t) \int_{-\infty}^t dt'
  \left[\varphi(t')^2 - \varphi(t)^2\right] \int \frac{d^3 {\bf p}}{(2 \pi)^3}
       {\rm Im}\left[ G_X^{++}({\bf p},t,t')\right]^2_{t>t'} \;,
\label{example}
\end{eqnarray}

\noindent
where $ G_X^{++}({\bf p},t,t')$ is the spatial {}Fourier transformed  causal
two-point propagator for the $X$ field.

The non-local term in the effective equation of motion represents a transfer
of energy from the $\varphi$ field into radiation, i.e. it is a dissipative
effect. Such terms can be localized when there is a separation of timescales
in the system (for discussion about the validity of the localization
approximation for the nonlocal term and region of parameters where this can be
achieved in specific models,
see e.g. Refs.~\cite{BMRnoise,BMR}).  Suppose, for example, that the
self-energy introduces a response timescale $\tau$. If $\varphi$ is slowly
varying on the response timescale $\tau$, 
\begin{equation}
\label{adiabatica}
\frac{\dot \varphi}{\varphi} \ll \tau^{-1},
\end{equation}
which is typically referred to as the adiabatic approximation, then we can use
a simple Taylor expansion and write 
\begin{equation}
f(\varphi(t')) - f(\varphi(t)) =(t'-t)\, \dot{\varphi}(t) \frac{\partial
  f(\varphi)}{\partial \varphi} +\dots
\end{equation}
The effective equation of motion for $\varphi$ (assumed homogeneous)
including the linear dissipative terms then becomes

\begin{equation}
\ddot\varphi+\Upsilon\dot\varphi+\frac{\partial V_{\rm
    eff.r}(\varphi)}{\partial\varphi}=0\;,
\label{eomdiss1}
\end{equation}
where $V_{\rm eff,r}(\varphi)$ is the renormalized effective potential and
$\Upsilon$
is the dissipation coefficient defined as
\begin{equation}
\Upsilon=\int d^4 x' \,\Sigma_R(x,x')\,(t'-t)
\label{cf},
\end{equation}
where $\Sigma_R(x,x')$ is a retarded correlation that depends on the specific
form  of the interaction $V_{\rm int}(\varphi, X)$ and is defined by

\begin{equation}
\Sigma_R (x,x') = - i \left[ \frac{\partial f(\varphi)}{\partial \varphi}
  \right]^2 \theta(t-t') \langle [ g(X(x)),g(X(x')) ] \rangle\;.
\label{Sigmacorr0}
\end{equation}

\section{Interaction Terms}
\label{interactions}

In this work we are interested in examining the case where a scalar
field $\phi$ has a background value $\langle \phi \rangle =
\varphi(t)$.  This field could be,  for example, the order parameter
in some phase transition or the inflaton field during cosmic inflation.  
{}For the purposes of these
calculations, the scalar field $\phi$ is treated generically.  This
field is coupled to other boson $\chi$ and fermion $\psi_{\chi}$
fields, which will be called the ``catalyst fields''.  These fields
are in turn coupled to bosonic $\sigma$ and fermionic $\psi_{\sigma}$
fields, which will be called the ``radiation fields''.  We will
compute the dissipation coefficients for the interaction
structures  listed below.  The combination of all the interaction
types to be considered encompass most of the couplings between scalars
and fermions that emerge in particle physics models.   In particular,
they can be combined with appropriately chosen coupling constants to
be applicable in Supersymmetric (SUSY) models, such as often used in
cosmological warm inflation model building
\cite{BasteroGil:2006vr,BasteroGil:2009ec}.  Alternatively they can be
used independently. In Appendix \ref{models} we will give some
examples. 

Next we specify the different types of interactions we will consider.
The interactions between catalyst and radiation fields are:

\begin{enumerate}

\item Interaction between the scalars $\chi$ and $\sigma$:

\begin{equation}
{\cal L}_I [\chi,\chi^\dagger,\sigma, \sigma^\dagger] = - h_1 M [\chi^\dagger
  \sigma^2 + \chi (\sigma^\dagger)^2 ] - g_3^2 \chi^\dagger\chi
\sigma^\dagger\sigma \;,
\label{Lchisigma}
\end{equation}
where $h_1$ and $g_3$ are coupling constants and $M$ is a mass term (in SUSY
models this often is $M=g_1 \varphi/\sqrt{2}$, where $\varphi$ is the
background inflaton field).

\item Yukawa interaction between the scalar $\chi$ and 
fermions $\psi_\sigma$,
$\bar{\psi}_\sigma$:

\begin{equation}
{\cal L}_I [\chi,\chi^\dagger,\psi_\sigma, \bar{\psi}_\sigma] = - h_2
[\chi^\dagger \bar{\psi}_\sigma P_R \psi_\sigma  
+ \chi \bar{\psi}_\sigma P_{L}
  \psi_\sigma  ]\;,
\label{Lchipsisigma}
\end{equation}
where $P_{R({L})}= (1 \pm \gamma_5)/2$ are the chiral projection operators.

\item Yukawa interaction between the scalar $\sigma$ and fermions
$\psi_\sigma, \bar{\psi}_\sigma, \psi_\chi,
\bar{\psi}_\chi$:

\begin{equation}
{\cal L}_I [\sigma,\sigma^\dagger,\psi_\sigma, \bar{\psi}_\sigma,
\psi_\chi, \bar{\psi}_\chi] = - h_3
[\sigma^\dagger \bar{\psi}_\chi P_R \psi_\sigma  
+ \sigma \bar{\psi}_\sigma P_{L}
  \psi_\chi  ]\;.
\label{Lchipsichipsisigma}
\end{equation}
 
\end{enumerate}

The interactions between the scalar $\phi$ field and the other fields are:

\begin{enumerate}

\item Interaction between the scalars $\phi$ and $\chi$:

\begin{equation}
{\cal L}_I [\phi, \phi^\dagger,\chi,\chi^\dagger] = - g_1^2
\phi^\dagger \phi \chi^\dagger\chi\;,
\label{Lphichi}
\end{equation}
where, without loss of generality, the background scalar field is
associated with the zero mode of the real component\footnote{Note that
  here we define the complex scalar fields in terms of their real and
  imaginary components like for example: $\phi = (\phi_1 + i
  \phi_2)/\sqrt{2}$ and similarly for $\chi$ and $\sigma$.} of $\phi$:
$\phi = (\phi_1 + \varphi + i \phi_2)/\sqrt{2}$.  {}From
Eq. (\ref{Lphichi}), the relevant term that will contribute to
dissipation in the scalar fields effective equation of motion, at
one-loop order, is then $- g_1^2 \varphi^2 | \chi|^2/2$.

\item Interaction between the scalar $\phi$ and the fermions $\psi_\chi$,
  $\bar{\psi}_\chi$,

\begin{equation}
{\cal L}_I [\phi,\phi^\dagger,\psi_\chi, \bar{\psi}_\chi] = - g_2
[\phi^\dagger \bar{\psi}_\chi P_R \psi_\chi  + \phi \bar{\psi}_\chi P_L
  \psi_\chi ]\;,
\label{Lphipsichi}
\end{equation}
where, in this case, the relevant term that will contribute to the scalar
field's effective equation of motion, is $- g_2 (\varphi/\sqrt{2})
\bar{\psi}_\chi \psi_\chi $.

\item Interaction among the three scalars $\phi$, $\chi$ and $\sigma$:

\begin{equation}
{\cal L}_I [\phi, \phi^\dagger,\chi,\chi^\dagger,\sigma, \sigma^\dagger] = -
g_1 h_1 [\phi^\dagger \chi^\dagger \sigma^2 + \phi \chi (\sigma^\dagger)^2
]\;.
\label{Lphichisigma}
\end{equation}

Note that the interaction (\ref{Lchisigma}) is the term dependent on the
scalar background field, derived from (\ref{Lphichisigma}) once $\phi$ is
written in terms of $\varphi$ and $M$ is then associated with $g_1
\varphi/\sqrt{2}$. The interaction term (\ref{Lphichisigma}) only contributes
to the dissipation in the  scalar field's effective equation of motion at
two-loop order.

\end{enumerate}

{}Finally there can be self-interactions for the scalars:

\begin{equation}
{\cal L}_I [\phi, \phi^\dagger] + {\cal L}_I [\chi,\chi^\dagger] + {\cal L}_I
[\sigma,\sigma^\dagger] = \lambda_\phi (\phi \phi^\dagger)^2 + \lambda_\chi
(\chi \chi^\dagger)^2 + \lambda_\sigma (\sigma \sigma^\dagger)^2\;.
\label{selfint}
\end{equation}

In the interaction forms considered above, at tree-level $\phi$ is
only coupled to either the scalar boson $\chi$ or fermion
$\psi_\chi$. In the presence of a vacuum expectation value  $\varphi$,
these can become heavier than the scalar $\sigma$ and fermion
$\psi_\sigma$, which in turn are only coupled at tree-level with the
intermediate fields $\chi$ and $\psi_\chi$.  At the level of model
interactions we are considering,  there are radiative (loop generated)
terms that can couple $\phi$ to $\sigma$ and $\psi_\sigma$, which can
make them also heavy fields. These radiative generated interactions
can however be kept small with either suitable choice of coupling
constants or by imposing a symmetry for the model interactions (like
SUSY).  These possibilities, together with physical examples, are
discussed in Appendix \ref{models}.  In the following sections we will always
assume that the radiation scalar $\sigma$ and fermion $\psi_\sigma$
fields are lighter than  the intermediate scalar boson $\chi$ or
fermion $\psi_\chi$ fields; it is left as a model building question
what conditions are required to achieve these conditions, but aside
from some comments in Appendix \ref{models}, those issues are beyond the subject
of this paper.

\subsection{Model Parameters}
\label{parameters}

The calculation will be done for different thermal regimes depending on the
relation between temperature $T$ and the field masses
$m_{\chi},m_{\psi_{\chi}},m_{\sigma},m_{\psi_{\sigma}}$, for the $\chi$,
$\psi_{\chi}$, $\sigma$ and $\psi_{\sigma}$ fields respectively:

\begin{itemize}

\item  Very low-temperature regime:
  $m_{\chi},m_{\psi_{\chi}},m_{\sigma},m_{\psi_{\sigma}} \gg T$.  

\item Low-temperature regime: 
$m_{\chi},m_{\psi_{\chi}} \gg T \gg m_{\sigma},m_{\psi_{\sigma}}$.

\item High-temperature regime:
  $m_{\chi},m_{\psi_{\chi}},m_{\sigma},m_{\psi_{\sigma}} \ll T$.

\end{itemize}
The very low-temperature regime has all field excitations suppressed
by Boltzmann exponential factors, and thus so are the dissipation
coefficients.  In the low-temperature regime the
interactions produce the two-stage catalyst induced dissipation of
\cite{BR1}.  This mechanism leads to dissipation 
that are not damped by Boltzmann exponential factors, despite the fact
the $\chi$ and $\psi_{\chi}$ fields are heavy with respect to $T$.
For the dissipative coefficients, the interactions that give the leading
$T$ dependent terms have been calculated in \cite{mossxiong0}, but here
we also present interactions that give less dominant terms
with respect to $T$.

Dissipation coefficients have already been computed  for
the self-interacting scalar field, in particular in
the high-temperature regime \cite{hosoya1,Masa}.  In this case,
for example in the computation of dissipation coefficients in the
local approximation, it demands much more stringent restrictions on
the model parameters \cite{BGR,BR1,BMRnoise} than dissipation coming
from $\phi$ interacting with other fields.  We will not be interested
in the dissipation effects from $\phi$  self-interaction. {}As such we
will assume the $\phi$ self-coupling $\lambda_\phi=0$, or we can
equivalently assume that $\phi$ is only a classical background field
which is coupled to the remaining fields (considered as the bath or
environment which is coupled to the system background field).

Concerning the magnitude of the many interaction terms considered
above, we will restrict to values such that the leading contributions
to the vacuum polarization terms (from which the decay terms are
extracted and, as we will see below, enter in the calculations of the
Green's functions) will come dominantly from one-loop processes, when the
decaying field is a heavy field (e.g. either $\chi$ or $\psi_\chi$).
{}For instance, we will keep the leading  one-loop order vacuum
polarization terms for the catalyst  (intermediate) fields $\chi$ and
$\psi_\chi$ coming from the radiation fields $\sigma$ and
$\psi_\sigma$, and that leads to decay processes like  $\chi \to
\sigma +\sigma$, $\chi \to \psi_\sigma + \bar{\psi}_\sigma$, and
$\psi_\chi \to \sigma + \bar{\psi}_\sigma$, with complex conjugate
analogous process.  Decay processes for the light boson and fermion
fields can only happen through Landau damping and, therefore vanish at
zero temperature  when on-shell. Decays for the heavy fields involving
two-loop terms, like Landau damping  coming from the fields
self-interactions or the bi-quadratic interaction will be considered
subdominant with respect the one-loop order processes.  {}For this to
be valid, the coupling constants in the interactions
(\ref{Lchisigma}), (\ref{Lchipsisigma}), (\ref{Lchipsichipsisigma})
and (\ref{selfint}) should satisfy

\begin{eqnarray}
&& g_3^2 \ll h_1, h_2 \nonumber \\ && \lambda_\chi \ll h_1, h_2\,.
\label{conditions1}
\end{eqnarray}
On the other hand, decay processes involving one light and one heavy
field are suppressed compared to the two-loop process involving the
$\sigma$ self-interaction term with coupling $\lambda_\sigma$.  The
only restriction we impose on the $\sigma$ self-interaction magnitude
is it to be small compared with the couplings, which then implies
in the constraint

\begin{equation}
\lambda_\sigma \ll g_3^2,h_2^2,h_3^2 \;,
\label{conditions2}
\end{equation}
so that temperature corrections  to the heavy fields
tree-level masses are larger than the temperature correction for the
$\sigma$. This way we can also keep the effective (temperature
dependent) masses for $\chi$ and  $\psi_\chi$ still larger than that
for the $\sigma$ and $\psi_\sigma$ in the high-temperature regime.

In the low-temperature regime, the above generic perturbative
constraints are sufficient for consistent calculations at one-loop
order with respect to the intermediate fields.  In the
high-temperature regime, however, computation of the dissipation 
coefficients are known to have issues regarding
resummation of higher loop terms \cite{jeon1,jeon,arnold}.  We will
return later on to the issue of  resummation when computing the
dissipation coefficients in the high-temperature
regime. There we will present some additional constraints on the model
parameters, in addition to (\ref{conditions1}),  so as to make the
calculation consistent in that regime.

\section{Dissipation Coefficients from Scalar and Fermion Interactions}
\label{dissvis2stage}

This section presents the specific dissipation  coefficients
emerging from the interaction terms given in Sect. \ref{interactions} and
that appear in the $\phi$-field background effective equation of
motion.

As discussed in Sec. \ref{diss}, the effective dynamics for the
background scalar field  when expressed in a local form is described by a
dissipative equation of motion of the form Eq. (\ref{eomdiss1}), with
dissipation coefficient $\Upsilon$ expressed in general by Eqs. (\ref{cf}) and
(\ref{Sigmacorr0}).

As mentioned before, we will assume  throughout that $\phi$ contributes only
through its  classical background field value.  
Moreover, these calculations will also be done in the adiabatic
approximation Eq. (\ref{adiabatica}) as well as 

\begin{equation}
\label{adiabatict}
\frac{\dot T}{T} \ll \tau^{-1} .
\label{adiabatica2}
\end{equation}

\begin{figure}[htb]
  \vspace{0.5cm} \epsfig{figure=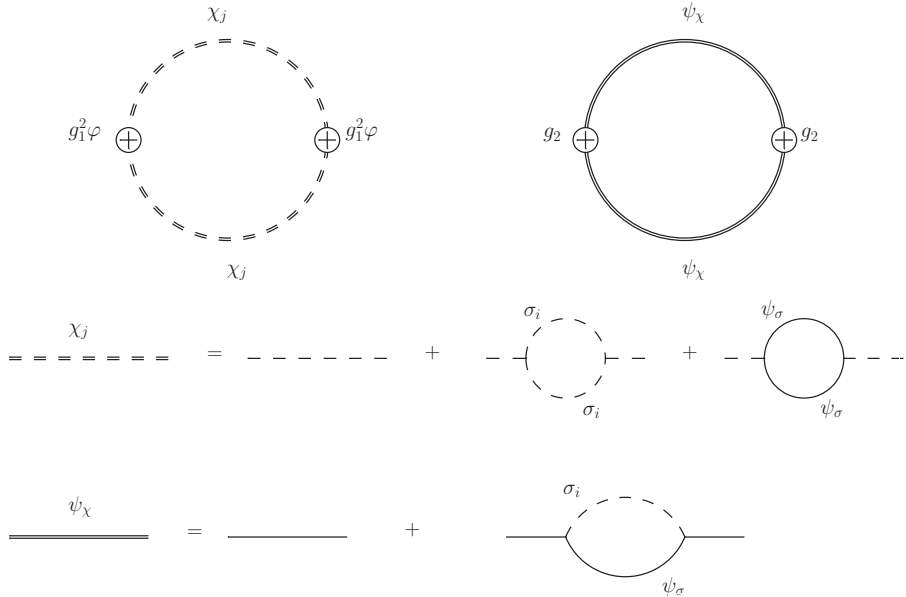,angle=0,width=12cm}
\caption[]{\label{chichi} The contribution to the dissipation coefficient at
  leading order. The thick propagator lines are dressed scalar $\chi$ (dashed)
  or fermionic $\psi_\chi$ (solid) propagators (shown here including the
  relevant contributions to vacuum polarization at one-loop order).}
\end{figure}

Thus, regarding
the interaction terms of the  intermediate (catalyst) fields  $\chi$ and
$\psi_\chi$ presented in Sect. \ref{interactions},  we find that $\Sigma_R$,
the retarded correlation entering in the definition of the dissipation
coefficient $\Upsilon$, Eq. (\ref{cf}), becomes

\begin{equation}
\Sigma_R (x') = -i \left(\frac{g_1^2}{2}\right)^2 \varphi^2 \sum_{i=1}^2
\theta(t-t')  \langle [ \chi_i^2(x'), \chi_i^2(0)] \rangle  -i \frac{g_2^2}{2}
\theta(t-t') {\rm tr} \left\{\langle [ \bar{\psi}_\chi(x') \psi_\chi(x'),
  \bar{\psi}_\chi(0) \psi_\chi(0) ] \rangle \right\}\;,
\label{Sigmacorr}
\end{equation}
where the first term in the above expression comes from the interaction
(\ref{Lphichi}) and the second term from (\ref{Lphipsichi}).  The dissipation
coefficient $\Upsilon$ as defined here is derived for a system near thermal
equilibrium.  As such, the averages in Eq. (\ref{Sigmacorr}) are over a
thermal equilibrium distribution.  Equation (\ref{Sigmacorr}) is
represented diagrammatically by {}Fig. \ref{chichi}, where only the relevant
contributions to the vacuum polarization (as discussed in the previous
section), which enter in the computation of the field propagators, are shown.

In terms of the interactions given by Eqs. (\ref{Lphichi}), (\ref{Lphipsichi})
and  (\ref{Lphichisigma}), the dissipation coefficient at
leading order is \cite{BMR},

\begin{eqnarray}
\Upsilon &=& \frac{2}{T} \left(\frac{g_1^2}{2} \right)^2 \varphi^2 \int
\frac{d^4 p}{(2 \pi)^4} \left[ \rho_{\chi_1}(p_0, {\bf p})^2 +
  \rho_{\chi_2}(p_0, {\bf p})^2 \right] n_{B}(p_0) \left[1 + n_{B}(p_0) \right]
\nonumber \\ &+& \frac{g_2^2}{2T} \int \frac{d^4 p}{(2 \pi)^4} {\rm tr} \left[
  \rho_{\psi_\chi}(p_0, {\bf p})^2 \right] n_{F}(p_0) \left[1 - n_{F}(p_0)
  \right]\;,
\label{Upsilon2}
\end{eqnarray}
where $\rho_\chi$ and $\rho_{\psi_\chi}$ are the spectral functions  for the
intermediate fields $\chi$ and $\psi_\chi$, defined 
in Sect. \ref{specfunc} by
Eqs.~(\ref{rhochi}) and (\ref{rhopsichi}), and $n_{B}$ and $n_F$ are the
Bose-Einstein ($n_{B}(\omega) = (e^{\beta \omega} - 1)^{-1}$)
and the Fermi-Dirac ($n_{F}(\omega) = (e^{\beta \omega} + 1)^{-1}$)
distributions, respectively, with $\beta \equiv 1/T$.

\section{Particle production interpretation}
\label{partprod}

\begin{figure}[htb]
  \vspace{0.5cm} \epsfig{figure=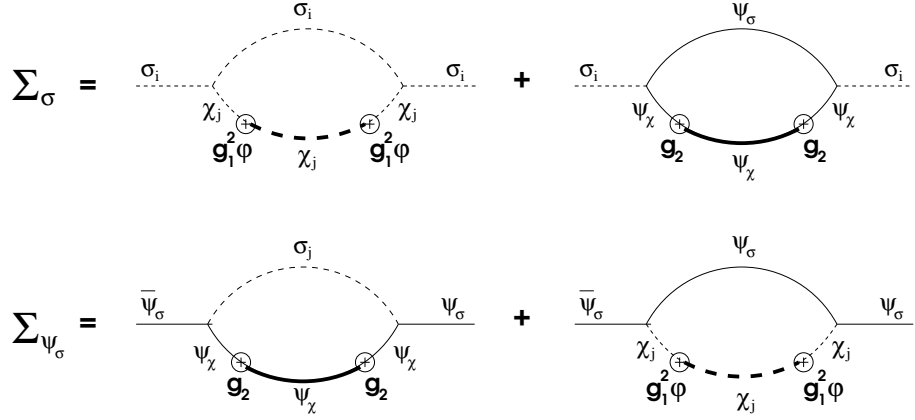,angle=0,width=12cm}
\caption[]{\label{selffigs} Diagrams contributing to the self-energy of the
  radiation bath fields.}
\end{figure}

The physics involving the appearance of dissipation terms in the background
effective equation of motion can be given an interpretation in
terms of radiation (or particle) production.  Such a physical
interpretation for the dissipation term (\ref{cf}) has been given
before~\cite{BR21,BR22,BMR,Graham}. We here briefly review this interpretation in terms
of the field interactions we are considering. Here dissipation can be seen as a
result of energy being transferred from the $\phi$ field to the radiation bath
fields $\sigma$ and $\psi_\sigma$ through the excitation of an intermediate
field, which in this case are either the scalars $\chi$ or fermions
$\psi_\chi$.  That this is indeed associated to particle production  ($\sigma$
and/or $\psi_\sigma$) has been demonstrated  explicitly in~\cite{Graham}. In
this case, the  particle production rate for the radiation fields are only due
to the interactions of $\chi,\psi_\chi$ with $\sigma,\psi_\sigma$, since 
the latter
do not  couple directly to the background field, except by the interaction
term Eq. (\ref{Lphichisigma}), which is however higher order  (two-loop), in
comparison to the other interactions.  The particle production rate of the
radiation bath  particles due to interactions is~\cite{Graham}

\begin{equation}
\dot{n} = {\rm Im} \left[ 2 \int_{-\infty}^t dt' \frac{e^{-i \omega({\bf
        p})(t-t')}}{2 \omega({\bf p})} \Sigma_{21} ({\bf p},t,t')\right]\;,
\label{dotn}
\end{equation}
where $\omega({\bf p})$ is the particle dispersion relation for the scalars
$\sigma$ or fermions $\psi_\sigma$ and
$\Sigma_{21}=\Sigma_{\sigma,21}+\Sigma_{\psi_\sigma,21}$ is the sum of the
$\sigma$ and $\psi_\sigma$ self-energies (in the Schwinger-Keldshy formalism
representation).  $\dot{n} = \dot{n}_\sigma+\dot{n}_{\psi_\sigma}$ is the
total particle production rate of the radiation bath particles
$\sigma,\psi_\sigma,\bar{\psi}_\sigma$ from the decay processes $\chi \to 2
\sigma$, $\chi \to \bar{\psi}_\sigma + \psi_\sigma$ and $\psi_\chi \to  \sigma
+\bar{\psi}_\sigma$.  The self-energies $\Sigma_{\sigma,21}$ and
$\Sigma_{\psi_\sigma,21}$ are  obtained (or vice-versa) from the diagrams
contributing to dissipation shown in {}Fig. \ref{chichi}, by cutting one of
the internal loop (joining the external) lines for $\sigma$  (in the case of
the $\sigma$ self-energy) or $\psi_\sigma$ (for the $\psi_\sigma$
self-energy). The diagrams representing these self-energies terms are shown in
{}Fig.~\ref{selffigs}.  An explicit expression for the self-energy is given
for example in~\cite{Graham}.

Using the expression of the total radiation energy,

\begin{equation}
\rho_{r} = \int \frac{d^3 p}{(2 \pi)^3} \omega({\bf p}) n\;,
\label{rhor}
\end{equation}
the dissipation coefficient Eq.~(\ref{cf}) can be expressed as~\cite{Graham}

\begin{equation}
\Upsilon = \frac{\dot{\rho}_r}{\dot{\varphi}^2} =
\frac{1}{\dot{\varphi}^2}\int \frac{d^3 p}{(2 \pi)^3}  \left[\omega_\sigma({\bf
    p}) \dot{n}_\sigma + \omega_{\psi_\sigma}({\bf p})
  \dot{n}_{\psi_\sigma}\right]\;.
\label{upsilon_n}
\end{equation}
This expression directly links the particle production of the bath radiation
fields with how the background field dissipates its energy, represented by the
dissipation coefficient $\Upsilon$. 

\section{Spectral Functions}
\label{specfunc}

Dissipation coefficients require the spectral functions for the
scalar and the fermion fields.
{}For the scalars (where $j=\chi_1,\chi_2,\sigma_1,\sigma_2$
denote the real components of the complex scalar fields) we have:

\begin{eqnarray}
\rho_{j} (p_0, {\bf p}) &=& \frac{i}{-p_0^2 + {\bf p}^2 + m_{R,j}^2
  + i {\rm Im} \Sigma_{j}(p)} -  \frac{i}{-p_0^2 + {\bf p}^2  +
  m_{R,j}^2 - i {\rm Im} \Sigma_{j}(p)} \nonumber \\ &=& \frac{4
  \omega_{j}({\bf p}) \Gamma_{j}(p_0,{\bf p})}{\left[-p_0^2 +
    \omega_{j}({\bf p})^2 \right]^2 + \left[ 2 \omega_{j}({\bf p})
    \Gamma_{j}(p_0,{\bf p}) \right]^2 }\;,
\label{rhochi}
\end{eqnarray}
where $\omega_{j}({\bf p})^2 = {\bf p}^2 + m_{R,j}^2$, with $m_{R,j}$
the effective, renormalized mass for the scalar fields, 
$m_{R,j}^2 =  m_{j}^2 + {\rm Re} \Sigma_{j}$.
Also in
Eq. (\ref{rhochi}) the decay width, $\Gamma_{j}(p_0,{\bf p})$, is defined
in terms of the imaginary part of the self-energy as 

\begin{equation}
\Gamma_{j}(p_0,{\bf p}) = \frac{{\rm Im} \Sigma_{j}(p)}{2
  \omega_{j}({\bf p})}\;.
\label{Gammachi}
\end{equation}
Note in the adiabatic approximation Eqs. (\ref{adiabatica}) and
(\ref{adiabatict}), which underlies all calculations here for dissipation 
coefficients, the response time scale is typically associated with
the decay widths, $\tau \rightarrow 1/\Gamma_j$.

{}For the fermions (where now $j=\psi_\chi,\psi_\sigma$) we have that

\begin{eqnarray}
\rho_{j} (p) &=& \frac{i}{ \not\!p + m_{R,j} + i {\rm Im}
  \Sigma_{j}(p) } - \frac{i}{ \not\!p + m_{R,j} - i {\rm Im}
  \Sigma_{j}(p) }\;,
\label{rhopsichi}
\end{eqnarray}
where $m_{R,j}$ is
the effective, renormalized mass for the fermion fields, 
$m_{R,j} =  m_{j} + {\rm Re} \Sigma_{j}$.
The imaginary part of the fermion self-energy appearing
in Eq. (\ref{rhopsichi}) can be written, using
rotational invariance, as (${\bf 1}$ is the 4x4 identity matrix)

\begin{equation}
{\rm Im} \Sigma_{j}(p) = \gamma^0 p_0 \tilde{\alpha}_0  +  {\bf
  \gamma}.{\bf p} \tilde{\alpha}_p + m_{R,j} \tilde{\alpha}_m  {\bf 1}
\;,
\label{ImSigmapsi}
\end{equation}
where the expressions for $\tilde{\alpha}_0, \tilde{\alpha}_p$ and
$\tilde{\alpha}_m$ are explicitly given
in Appendix \ref{app}.

The imaginary part for the self-energies, appearing in
Eqs. (\ref{rhochi}) and (\ref{rhopsichi}) are derived in Appendix
\ref{app} for completeness.  The results for the decay widths are also
given in Appendix \ref{app}.  {}For example, for the bosonic
interaction Eq. (\ref{Lchisigma}), it is given by Eq. (\ref{Gamma1})
with $g_X^2= 4 (h_1 M/\sqrt{2})^2$, while for the Yukawa interaction
Eq. (\ref{Lchipsisigma}), it is given by Eq. (\ref{Gamma2}) with
coupling factors now defined by $g_X^2= (h_2/\sqrt{2})^2$, and for the
mixed boson-fermion interaction Eq. (\ref{Lchipsichipsisigma}), it is
obtained, likewise, from Eqs. (\ref{Gamma3}) - (\ref{alpham}), with
$g_X^2 = h_3^2$.

\section{Dissipation  Coefficients in the Low and High 
Temperature Regimes}
\label{results}

An analysis of Eq. (\ref{Upsilon2}) indicates that the behavior 
of the dissipation coefficients in the different temperature and interaction
regimes is determined by an interplay between the spectral function
and the thermal occupation numbers.  There are two regimes where well
defined approximations can be made. One is the regime of fixed $T$ and
where the couplings of the catalyst fields with the light fields,
$h_i$, are going to zero.  In this regime, the poles of the spectral
functions will dominate the integral and this will be referred to as
the {\it pole approximation}. The other regime is the integration region
around low energy $p_0$ and  low three-momentum ${\bf p}$, and  this
will be referred to as the {\it low-momentum approximation}.  In the pole
approximation, a well defined peak occurs in the spectral function, at
an energy around the heavy particle dispersion relation.  When the
temperature is large, the occupation numbers are also large, and this
approximation is valid. As such, the pole approximation works well in
the high-$T$ region.  In the low-momentum approximation,  and at low
temperatures, $T \leq m_\chi, m_{\psi_\chi}$, the spectral function
still has a well defined peak at around the heavy particle dispersion
relation, but now this contribution is exponentially suppressed due to
the occupation numbers in Eq.  (\ref{Upsilon2}).  
However even at low $T$ the occupation  numbers in the
low energy and momentum region will not be exponentially suppressed,
and so this approximation dominates over the pole contribution, which
is hugely suppressed from the exponentially damped occupation numbers.
Thus  at low $T$ the low-momentum approximation dominates. {}Finally,
at very low $T$ when also the decay widths become exponentially
damped, both the pole and low-momentum approximation can be important,
depending on the size of the couplings of the catalyst fields with the
light fields, $h_i$, and the masses of all the particles involved.  

Both approximations are now examined in detail. In the pole
approximation, valid for fixed $T$ and $h \rightarrow 0$, the general
expression for the dissipation coefficient 
Eq. (\ref{Upsilon2}) take on the
simplified form,

\begin{eqnarray}
\Upsilon &\simeq &  \frac{1}{T} \left(\frac{g_1^2}{2} \right)^2 \varphi^2
\sum_{i=1}^2 \int \frac{d^3 p}{(2 \pi)^3}
\frac{1}{\Gamma_{\chi_i}(\omega_{\chi_i}, {\bf p}) \omega_{\chi_i}^2}
n_B(\omega_{\chi_i}) \left[1 + n_B(\omega_{\chi_i}) \right] \nonumber \\ &+&
\frac{2g_2^2}{T} \int \frac{d^3 p}{(2 \pi)^3}
\frac{m_{R,\psi_\chi}^2}{\Gamma_{\psi_\chi}(\omega_{\psi_\chi}, {\bf p})
  \omega_{\psi_\chi}^2} n_F(\omega_{\psi_\chi}) \left[1 -
  n_F(\omega_{\psi_\chi}) \right]\;,
\label{UpsilonhighT}
\end{eqnarray}
where $\Gamma_{i}$ is the total decay width, e.g. for $\chi_{1,2}$ it
is  given by the sum of Eqs. (\ref{Gamma1}) and (\ref{Gamma2}),  while
for fermions, e.g. $\Gamma_{\psi_\chi}$, the fermion $\psi_\chi$ decay
width is defined by the imaginary part  of the pole of the fermion
spectral function Eq. (\ref{rhopsichi}) and is  given for example by
Eq. (\ref{Gamma3}).  

{}For the low-momentum approximation, valid for
fixed couplings $h_i$ and  $T \rightarrow 0$, the spectral functions
for the scalar boson fields becomes

\begin{equation}
\rho_{i} \simeq \frac{4}{m_{R,i}^3 }\Gamma_{i}\;,
\label{rhochilowT2}
\end{equation}
while the spectral function for the fermion
fields, Eq. (\ref{rhopsichi}), becomes

\begin{equation}
\rho_{i} \simeq \frac{2}{m_{R,i}^2 }\left( \tilde{\alpha}_0
\gamma_0 p_0 + \tilde{\alpha}_p  {\bf \gamma}.{\bf p}+ m_{R,i}
\tilde{\alpha}_m {\bf 1} \right) \;,
\label{rhopsichilowT2}
\end{equation}
and then

\begin{equation}
{\rm tr} (\rho_{i}^2) \simeq 16  \frac{(p_0^2 \tilde{\alpha}_0^2-{\bf
    p}^2\tilde{\alpha}_p^2  + m_{R,i}^2 \tilde{\alpha}_m^2 )
}{m_{R,i}^4}\;.
\label{trrhopsi2}
\end{equation}
In this approximation, the dissipation coefficient, Eq. (\ref{Upsilon2}),
using Eqs. (\ref{rhochilowT2}) and (\ref{trrhopsi2}), becomes

\begin{eqnarray}
\Upsilon &=& \frac{2}{T} \left(\frac{g_1^2}{2} \right)^2 \varphi^2 \sum_i \int
\frac{d^4 p}{(2 \pi)^4} \frac{16}{m_{R,i}^6 }\Gamma_{i}^2 \,n_{B}(p_0) 
\left[1 + n_{B}(p_0) \right]
\nonumber \\ 
&+& \frac{g_2^2}{2T} \int \frac{d^4 p}{(2 \pi)^4} 
16  \frac{(p_0^2 \tilde{\alpha}_0^2-{\bf
    p}^2\tilde{\alpha}_p^2  + m_{R,i}^2 \tilde{\alpha}_m^2 )
}{m_{R,i}^4} n_{F}(p_0) \left[1 - n_{F}(p_0)
  \right]\;.
\label{UpsilonlowT2}
\end{eqnarray}
Note from Eqs. (\ref{UpsilonhighT}) and (\ref{UpsilonlowT2}) that the dependence
of the dissipation coefficient on the heavy fields catalyst decay widths 
is very different in the two regimes where the pole and low temperature
approximations are applicable. While in the pole approximation the dissipation
coefficient is inversely proportional to $\Gamma_i$, in the low temperature
regime it is proportional to $\Gamma_i^2$. 
These different dependencies will be of fundamental
relevance when discussing higher order contributions to the dissipation as we will
analyze later on.
Next, we analyze the dissipation coefficient coming from the scalar and fermion
interactions in the many different energy regimes with respect to 
the field masses and the temperature.

\subsection{Very Low Temperature Regime}

In the very low-temperature regime,  $T \ll m_\sigma,m_{\psi_\sigma},
m_{\chi_i}, m_{\psi_\chi}$, the decay widths for the heavy
fields into light fields, which can be read from Eqs. (\ref{Gamma1}),
(\ref{Gamma2})  and (\ref{Gamma3}), with $X$ the heavy field
and the decay products $X_1,X_2$ corresponding to the light fields,
are dominated by the zero temperature contributions,

\begin{eqnarray}
&&\Gamma_{\chi}^{(1)} \approx \left(\frac{h_1M}{\sqrt{2}}\right)^2  
\frac{1}{ 8 \pi m_{\chi}}\;, \nonumber \\ 
&&\Gamma_{\chi}^{(2)}
  \approx \left(\frac{h_2}{\sqrt{2}}\right)^2 \, \frac{m_\chi}{ 8 \pi} \;, 
\nonumber \\ 
&& \Gamma_{\psi_\chi}
  \approx m_{\psi_\chi} \tilde{\alpha}_m \approx h_3^2  \,
  \frac{ m_{\psi_\sigma}}{ 4 \pi}\;,
\label{GammaslowT}
\end{eqnarray}
with temperature corrections that are Boltzmann suppressed by the heavy 
catalyst field masses.

Both the pole and low-momentum approximation have regions of validity
and in both cases, as we will see below, the dissipation 
coefficient  associated with a heavy field decay get exponentially
suppressed.  In general in this very low-temperature
regime, for small couplings $h_i$, the pole approximation,
Eq. (\ref{UpsilonhighT})
holds.  The integrals in this expression are dominated by  low
three-momentum with respect to the field masses, to give

\begin{eqnarray}
&&\Upsilon \stackrel{T\ll m, {\rm low}\; h}{\approx}  {\cal O}\left[
    \frac{g_1^2 T^2}{\Gamma_\chi^{(1)} + T^2\Gamma_\chi^{(2)}/m_\chi^2}
    e^{-m_\chi/T}\right]+ {\cal O}\left( \frac{g_2^2 T^2}{\Gamma_{\psi_\chi}}
  e^{-m_{\psi_\chi}/T}\right).
\label{UpsilonverylowTlowh} 
\end{eqnarray}
\noindent
{}For large values of couplings $h_i$ in this very low-temperature
regime, the dominant contributions to the dissipation
coefficient, Eq. (\ref{Upsilon2}),
come  from the low-momentum approximation to the spectral functions,
Eqs. (\ref{rhochilowT2}) and (\ref{rhopsichilowT2}), with $p_0, |{\bf
  p}| \ll  m_{\chi_i}, m_{\psi_\chi},m_{\sigma_i}, m_{\psi_\sigma}$.
Due to the Bose-Einstein and {}Fermi-Dirac distributions in
Eq. (\ref{Upsilon2}), the larger
contribution to the energy integrals now comes from $T \sim p_0$. The
dissipation coefficient can then be approximated as 

\begin{eqnarray}
&&\Upsilon \stackrel{T\ll m}{\approx}  
{\cal O}\left[ \frac{g_1^2
      T^3}{m_\chi^4} \left(\Gamma_\chi^{(1)} + T^2\Gamma_\chi^{(2)}/m_\chi^2
    \right)^2\right]+ {\cal O}\left( \frac{g_2^2 T^5}{m_\chi^4}
  \Gamma_{\psi_\chi}^2 \right)\;.
\label{UpsilonverylowT} 
\end{eqnarray}
The behavior seen above is directly proportional to the square of the decay
rates.  If $p_0 \gtrsim p$, the energy conditions in
Eqs.~(\ref{Gamma1}), (\ref{Gamma2}), (\ref{alpha0}), (\ref{alpham})
and (\ref{alphap}) cannot be satisfied and the   the dissipation 
coefficient vanishes identically,  $\Upsilon = 0$. 
{}For $p_0 \lesssim p$, only the Landau damping terms in
the  decay rates contribute. But these are exponentially (Boltzmann) suppressed,
thus the dissipation coefficients in this case are negligibly small.
As $T\to 0$,  $\Upsilon \to 0$, as it is expected~\cite{BMRnoise}.

\subsection{Low Temperature Regime}
\label{lowTsubsec}

The low-temperature regime is defined as
$m_{R,\sigma},m_{R,\psi_\sigma} \ll T \ll m_{R,\chi_i},
m_{R,\psi_\chi}$.  In other words, this is the regime of low
temperature with respect to the heavy catalyst fields but high
temperature with respect to the radiation fields.   In general the
masses of the radiation bath fields $m_\sigma,m_{\psi_\sigma}$ can be
neglected.  The temperature corrections to the effective masses of the
heavy fields, due to the light field self-energies, can also be
neglected, e.g.,  $m_{\rm R,\chi}^2 \simeq m_\chi^2 + (4 \lambda_\chi
+g_3^2 + h_2^2) T^2/12  \approx m_\chi^2$ and $m_{\rm R,\psi_\chi}^2
\simeq m_{\psi_\chi}^2 +  h_3^2 T^2/12 \approx m_{\psi_\chi}^2$,
likewise for the light fermion field $\psi_\sigma$, whose loop
radiative corrections, including the thermal ones, are suppressed by
the heavy field masses, even though in this case the light fields are
in the high-temperature regime.  However the light scalar $\sigma$
field effective mass will get a temperature correction from
self-interaction,  $m_{R,\sigma}^2 \simeq m_\sigma^2+\lambda_\sigma
T^2/3$.

In this regime one can verify that the dominant contributions to the
dissipation, Eq. (\ref{Upsilon2}), come from  $p_0, |{\bf p}| \ll m_{R,\chi_i},
m_{R,\psi_{\chi_i}}$,  which lead to the low-momentum approximation.
There is also a regime for fixed $T$ and $h \rightarrow 0$ in which
the pole approximation is valid.  
Since the temperature is small with
respect the heavy field masses, the three momentum integration in
Eq. (\ref{UpsilonhighT})
is dominated by   $|{\bf p}| \ll m_{R,\chi_i}, m_{R,\psi_{\chi_i}}$.
In the pole approximation the decay widths are just given by the zero
temperature results in Eqs. (\ref{Gamma1}), (\ref{Gamma2}) and
(\ref{Gamma3}).  The behavior of the dissipation 
coefficient is, however, exponentially suppressed and of the form of
Eq.  (\ref{UpsilonverylowTlowh}).

{}For the case of fixed $h$ with $T \rightarrow 0$, the low-momentum
approximation is valid, so the spectral functions can be approximated
as Eqs. (\ref{rhochilowT2}) and (\ref{rhopsichilowT2}).  The dissipation
coefficient is now given by Eq. (\ref{UpsilonlowT2}), which then 
leads in this low-temperature regime to the result:

\begin{eqnarray}
\Upsilon &\simeq & \frac{g_1^4 \varphi^2 T^3}{8 \pi^5} \sum_{i=1}^2
\frac{1}{m_{R,\chi_i}^6} \left[ A_1 \frac{1}{m_{R,\chi_i}^2} \left(\frac{h_1
    M}{\sqrt{2}} \right)^4 + B_1 \left( \frac{h_2^2 T^2}{m_{R,\chi_i}}
  \right)^2 + 2 C_1 \frac{h_2^2 T^2}{m_{R,\chi_i}^2} \left(\frac{h_1
    M}{\sqrt{2}} \right)^2 \right] 
+ D_1 \frac{g_2^2 h_3^4}{64
  \pi^5} \frac{T^5}{m_{R,\psi_\chi}^4}\;,
\label{UpsilonlowT}
\end{eqnarray}
where the constants in the above equation arise from evaluation of the
momentum and energy integrals in Eq. (\ref{UpsilonlowT2}). 
They have been evaluated numerically to give, 
$A_1\simeq 124.9$, $B_1 \simeq 129.8$, $C_1 \simeq 18.2$, $D_1 \simeq 4362.6$.  

Note from Eq. (\ref{UpsilonlowT}) that at leading order the dissipation coefficient
goes as ${\cal O}(T^3/m_\chi^2)$, agreeing with \cite{mossxiong0}.
The dissipation coefficient obtained in this low temperature and low momentum
regime is much less suppressed than what
would be expected from the pole approximation case.  
In particular, we obtain no exponentially
suppressed results as in the very low temperature regime. 

The heavy fermion loop contribution is worth noting in
Eq. (\ref{UpsilonlowT}).
It is given by the last term on the RHS of that expression.  
The fermion contribution is sub-leading in the temperature in the low-$T$
regime, as compared with the term coming from the decay into only
bosons.

\begin{figure}[htb]
  \vspace{1.1cm} \epsfig{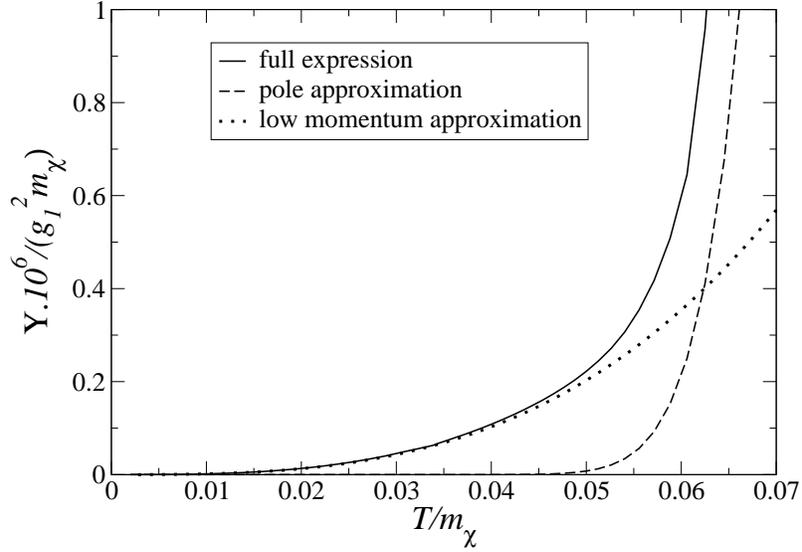}
\caption[]{\label{diss-T} The behavior of the dissipation coefficient as a
  function of the temperature (in the low-temperature regime), for $g_1=g_2$,
  $h=h_1=h_2=h_3=0.5$, $m_\sigma=0.001 m_\chi$ and $m_{\psi_\chi}=m_\chi$,
  $m_{\psi_\sigma}=m_\sigma$.}
\bigskip
\end{figure}

\begin{figure}[htb]
  \vspace{1.1cm} \epsfig{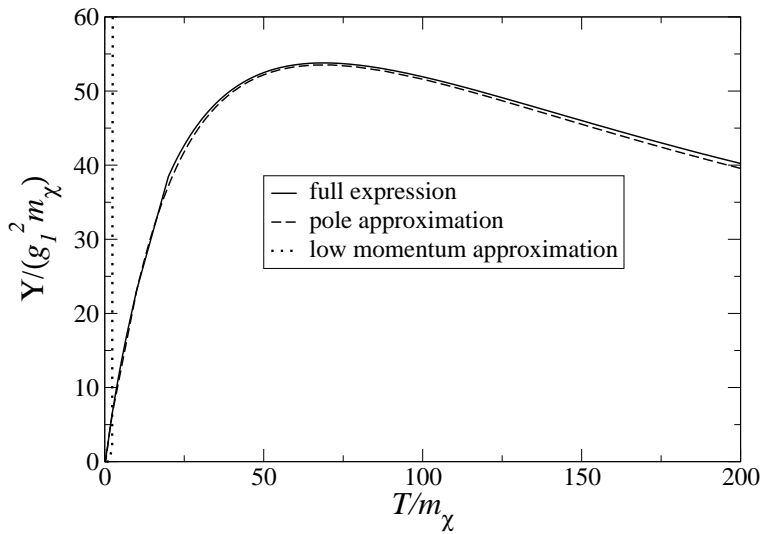}
\caption[]{\label{highT1} The behavior of the dissipation coefficient as a
  function of the temperature, for $g_1=g_2$, $h=h_1=h_2=h_3=0.5$, $g_3=0.01$,
  $m_\sigma=0.001 m_\chi$ and $m_{\psi_\chi}=m_\chi$,
  $m_{\psi_\sigma}=m_\sigma$.}
\end{figure}

This result can be explicitly verified by  numerically solving the
full expression, Eq.
(\ref{Upsilon2}),  and comparing with
the respective approximations in each temperature regime case. This is done  
in {}Figs. \ref{diss-T} and  \ref{highT1}, for
choice of coupling $h = 0.5$ (where we take for simplicity
$h_1=h_2=h_3=h$ in the expressions  for the decay widths) as function
of temperature $T$, at both low and high temperature respectively.
It is clear from {}Fig.  \ref{diss-T} that as $T \ll m_{\chi}$, the
low-momentum approximation agrees well with the full expression
Eq. (\ref{Upsilon2}), 
whereas the pole expression agrees poorly as
expected.  
The reliability of the results at high temperature, where the pole 
approximation is applicable, will be analyzed in
more details in the next subsection.

{}From these comparisons, we see in all cases, the low-momentum
approximation works well at very low temperatures $T/m_{\chi}
\stackrel{<}{\sim} 0.04$, whereas at higher temperatures $T/m_{\chi} >
1.0$, the pole approximation works well.  The breakdown of the
low-momentum approximation as $T$ increases arises because the
occupation numbers  in Eq. (\ref{Upsilon2})
no longer become exponentially suppressed, and so the
contribution from the poles start to dominate.  
Results for  the dissipation  coefficient, coming from the 
individual interaction types, in the low-temperature regime, are 
summarized in  Table \ref{tab1}.

\begin{table}
\begin{tabular}{c|c}
\hline
${\cal L}_I$ & dissipation $\Upsilon$ \\ 
\hline 
$-g_1^2 \phi^{\dagger} \phi \chi^{\dagger} \chi
  -h_1 M[\chi^\dagger \sigma^2 + \chi (\sigma^\dagger)^2 ]$ &  $0.026 g_1^4
  h_1^4 \varphi^2 {T^3 M^4}/{m_{R,\chi}^8}$  \\ 
\hline 
$-g_1^2
  \phi^{\dagger} \phi \chi^{\dagger} \chi - h_2[\chi^{\dagger} {\bar
      \psi}_{\sigma} P_R \psi_{\sigma} + \chi {\bar \psi}_{\sigma} P_L
    \psi_{\sigma}]$ & $0.11 g_1^4 h_2^4 \varphi^2 {T^7}/{m_{R,\chi}^8}$ \\ 
\hline 
$-g_1^2 \phi^{\dagger} \phi \chi^{\dagger} \chi -h_1
  M[\chi^\dagger \sigma^2 + \chi (\sigma^\dagger)^2 ]$ &  $0.015 g_1^4 h_1^2
  h_2^2 \varphi^2 {T^5 M^2}/{m_{R,\chi}^8}$ \\ 
$\hspace{-0.2cm} - h_2[\chi^{\dagger} {\bar \psi}_{\sigma} P_R
    \psi_{\sigma} + \chi {\bar \psi}_{\sigma} P_L \psi_{\sigma}]$ & \\ 
\hline $- \frac{1}{\sqrt{2}} g_2 \varphi {\bar \psi}_{\chi} \psi_{\chi}
  -h_3[\sigma^{\dagger} {\bar \psi}_{\chi} P_R \psi_{\sigma} + \sigma {\bar
      \psi}_{\chi} P_L \psi_{\sigma}]$ & $0.22 g_2^2 h_3^4
  T^5/m_{R,\psi_{\chi}}^4$ \\
\hline
\end{tabular}
\caption{A summary of all the dissipation  coefficients
in the low-temperature regime coming from each of the
interaction cases involving a heavy intermediate
field
(all expressions evaluated
with $m_{\chi_1} = m_{\chi_2}$).}
\label{tab1}
\end{table}

\subsection{High Temperature Regime}
\label{highTsec}

In this regime, we still have $m_\sigma,m_{\psi_\sigma}\ll T$.  
As for the heavy catalyst field contributions,
their effective thermal masses are $m_{\rm R,\chi_i}^2 \simeq
m_{\chi_i}^2 + (4 \lambda_\chi+g_3^2 + h_2^2) T^2/12$ and $m_{\rm
  R,\psi_\chi}^2 \simeq m_{\psi_\chi}^2 + h_3^2 T^2/12$.  Two
temperature regions can be useful to define, $ m_{\chi_i}, m_{\psi_\chi}
\ll T < \sqrt{12} m_{\chi_i}/\sqrt{4 \lambda_\chi +g_3^2 + h_2^2},
\sqrt{12} m_{\psi_\chi}/h_3$, and $\sqrt{12} m_{\chi_i}/\sqrt{g_3^2 +
  h_2^2},  \sqrt{12} m_{\psi_\chi}/h_3 < T$.  In both cases
Eq. (\ref{Upsilon2}) lead to a
high-temperature expression for dissipative 
coefficient, which is dominated by the pole approximation and lead
to the approximate expression Eq. (\ref{UpsilonhighT}). 
In this temperature regime, leading order one-loop expressions 
for these coefficients can become problematic as we will see below
and are only valid in a restricted region of parameter space.  
The reason for this can be seen from how the dissipative coefficient in
Eq. (\ref{UpsilonhighT})
depends on the field decay widths $\Gamma_i$.  Since it is 
inversely proportional to $\Gamma_i$ and since the $\Gamma_i$ are
proportional to the square of the coupling constants (see the results
in Appendix \ref{app}), loop corrections come with a ratio of the
vertex coupling squared to the  decay width coupling squared.
Depending of the magnitude of these two couplings, in some parameter
regimes higher loop correction terms  to the dissipation 
will be the same order as the one-loop term, thus
requiring resummation, and in other parameter regimes higher loop
contributions will be suppressed.

\subsubsection{Resummation issues}
\label{7C1}

It is worth recalling at this point the origin of this  dependence on
the couplings.  At high temperature, diagrams contributing to the
dissipation would have near on-shell
singularities if only products of bare propagators were used.  These
singularities are softened by introducing explicit lifetimes for
excitations, thus the decay widths $\Gamma_i$, through dressed
propagators.  This procedure leads to the result shown in
Eq. (\ref{UpsilonhighT})
with a similar procedure at higher loops.  The presence of these
regulating thermal widths  in the denominator then imply entire
classes of diagrams, called ladder diagrams, which are diagrams with
insertions  
of loops between the external propagators and in the definition of the
dissipation they can contribute at the same
order. This is much like what happens in the definition of
viscosity coefficients from Kubo formulas as seen for example in 
Refs. \cite{jeon,jeon1}, in the case of a single scalar
field, and full resummation schemes to account for all the corrections  
occurring at the same order
were used. Here we apply the outcome of this analysis for
our case of field interactions involving light radiation and heavier
catalyst fields.

\begin{figure}[htb]
  \vspace{0.5cm} \epsfig{figure=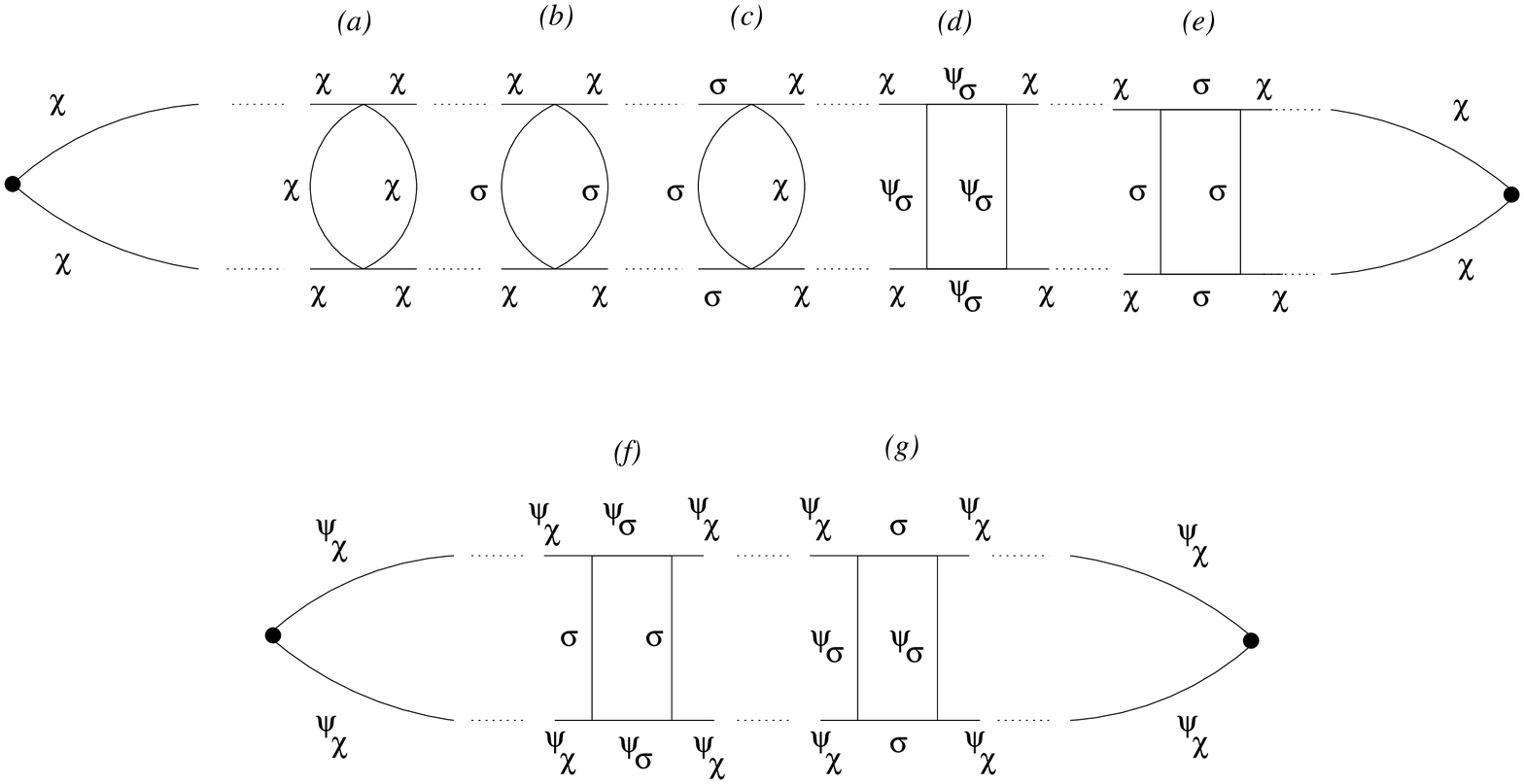,angle=0,width=12cm}
\caption[]{\label{multiloop} Examples of topologies of diagrams
that can contribute to the dissipation beyond the one-loop order.}
\end{figure}

Two possible examples of topologies of diagrams  that can
contribute beyond one-loop order are shown in
{}Fig. \ref{multiloop}. Higher order diagrams can be made  with
different variations of these topologies and similar ones.

Using the simple cutting rules explained e.g. in
Refs. \cite{jeon,jeon1}, let us first consider the contributions in the
first topology shown by the top diagram in
{}Fig. \ref{multiloop}, which are the types of topologies contributing to dissipation
due to the heavy catalyst scalar field $\chi$. Higher order contributions involve
adding rungs to the simple leading order contribution, which are of the form
of the rungs (a)-(e). The analysis from these contributions to dissipation shows
propagators which can share pinching pole singularities. These are propagators for the
fields running on the external part of the diagram that can share the same momentum
when going on-shell, thus displaying near on-shell pinching poles. As mentioned
above, these singularities are regulated by the field's decay width,
giving as a leading order contribution to the dissipation
for instance a contribution of order $(g_{i}^2/\Gamma_i)^n$, depending the type
and number of rungs added to the diagram.  {}For example, take
for instance the case of a rung of type (a) shown in the top topology in {}Fig. \ref{multiloop}. 
In this case the contribution
of the rung (a) produces a multiplicative contribution to the leading order
one-loop dissipation  term that is ${\cal O}(\lambda_\chi^2/\Gamma_\chi)$.
$\Gamma_\chi$ receives contributions coming from the $\chi$
self-coupling vertex, $\sim {\cal O}(\lambda_\chi^2)$, from the
coupling with the light $\sigma$  bosons, $\sim {\cal O}(h_1^2)$, and
from light $\psi_\sigma$ fermions,  $\sim {\cal O}(h_2^2)$. In this
case, this contribution can be rendered small compared to the one-loop
result for the heavy field provided that $h_1,h_2 \gg
\lambda_\chi$. Another possible contribution comes from rungs of type (b).  
Its contribution is ${\cal O}(g_3^4/\Gamma_\chi)$, which can be rendered small by
the same considerations taken for type (a). The next case are contributions involving
rungs like type (c). Its contribution is now of  ${\cal O}(g_3^4/\Gamma_\sigma)$
The thermal width for $\sigma$ comes from the imaginary part of the
self-energy terms for $\sigma$ built with the interactions (\ref{Lchisigma}),
(\ref{Lchipsichipsisigma}), (\ref{Lphichisigma}) and its self-interaction term
in (\ref{selfint}). The leading order contributions from these interactions
give $\Gamma_\sigma \sim {\cal O}(\lambda_\sigma^2)+{\cal O}(h_1^2)+{\cal O}(h_3^2)+
{\cal O}(g_3^4) \sim {\cal O}(h_1^2,h_3^2)$. These contributions can also be rendered small,
compared with the one-loop contribution with external propagators
given by the heavy field $\chi$, with a suitable choice of couplings,
e.g. $g_3 \ll h_1,h_2,h_3$.  The next two cases of contributions
involves rungs of the form (d) and (e). The contributions of the form (d)
gives ${\cal O}(h_2^4/(\Gamma_\chi\Gamma_{\psi_\sigma}))$. Since
$\Gamma_{\psi_\sigma}\sim {\cal O} (h_2^2 + h_3^2)$ and using e.g. the typical
parameters found in SUSY models displaying the types of interactions we have,
the couplings will satisfy Eq. (\ref{susycouplings}), and we find that these
contributions can potentially be of ${\cal O}(1)$. Likewise, the contribution
of type (e) for the couplings shown in Eq. (\ref{susycouplings}) gives a parametric
result of ${\cal O}(g^4/h^2)$, which is higher order. 
The last topologies to be analyzed are the ones contributing to the dissipation coming
from heavy catalyst fermion field and given by diagrams of the form of the 
bottom diagram shown in  {}Fig. \ref{multiloop}. The relevant additional rungs that 
appear are of the form of the terms (f) and (g) shown in that diagram.
The contribution from rungs of the form (f) are parametrically  
${\cal O}(h_3^4/(\Gamma_{\psi_\chi}\Gamma_{\psi_\sigma}))$. Using that 
$\Gamma_{\psi_\chi}\sim {\cal O}(h_3^2)$, and parameters shown in Eq. (\ref{susycouplings}),
we find that this contribution is potentially of ${\cal O}(1)$. The other contribution 
comes from rungs of the form (g), which gives ${\cal O}(h_3^4/(\Gamma_{\psi_\chi}\Gamma_{\sigma}))
\sim {\cal O}(1)$ using again Eq. (\ref{susycouplings}). This contribution is also 
potentially of same order as the one-loop result for the dissipation due
to heavy fermions in the higher temperature regime. The fermion dissipation, however, 
tend to be numerically smaller than the scalar field contributions to
dissipation at high temperatures, as we will see from the results to be shown below and from
those in {}Fig. \ref{diss-ones}.

\subsubsection{One-loop results}

Though in principle all higher loop contributions to the dissipation 
can be re-summed using a variety of techniques, we will not attempt to do
this here.  The interactions considered in this paper  are much more
complex than the simple case of only a single scalar field, which
was considered in previous works performing these resummations,
and it would require a
separate and thorough analysis to examine the necessary resummation.
In any event, it is important to recognize that in contrast to the
original analysis made e.g. in Refs. \cite{jeon1,jeon} that had only a single
coupling parameter, in our case where we have  multiple coupling
parameters and, as discussed above, we now have the added freedom to choose
certain parameter regimes where some of higher loop contributions can be
suppressed and no resummation for them are in principle necessary.  This is true
for example when computing particular contributions of fields, 
where for example, for those classes of
diagrams that involve intermediate heavy field propagators and exemplified in
the previous subsection.  The
cases that need special attention with respect to resummation are those where 
light field components run in the external part of the diagrams contributing
to dissipation. This is e.g. the cases involving the inclusion of rungs of
type (c) and (e) in the top topology shown in {}Fig. \ref{multiloop},
or the rungs (f) and (g) in the second topology of the same figure, 
and seen in the parametric analysis performed  previously. 

Below we show some results for the dissipation at one-loop order at high temperatures, 
but these results are only meant to be phenomenological ones, 
providing an estimate for the dissipation, since they do not account for all the
possible higher order loop contributions that they may receive.      

In {}Fig. \ref{highT1} are shown the result for the total dissipation coefficient,
for the case of summing all the contributions,
for temperatures up to two hundred times the intermediate field mass.
Note that the low-momentum approximation quickly
overestimates the true result as the temperature increases.  The pole
approximation becomes very good  already at  $T \sim m_{\chi},
m_{\psi_\chi}$, but tends to slightly lose accuracy at higher
temperatures, where the
simple pole approximation, given in terms of the particles dispersion
relation, becomes less accurate.  This is more critical in the case
when only decay into bosons are considered (e.g. with decay width
given by Eq. (\ref{Gamma1})), and the temperature is such that
$\Gamma_\chi^{(1)}/m_{\chi} \gtrsim 1$ (note that this cannot happen
in the case of decay into fermions, where for large temperatures the
decay width tends to zero (on-shell) due to Pauli blocking
\cite{boya1}).  
 
\begin{figure}[htb]
\bigskip
  \vspace{1.1cm} \epsfig{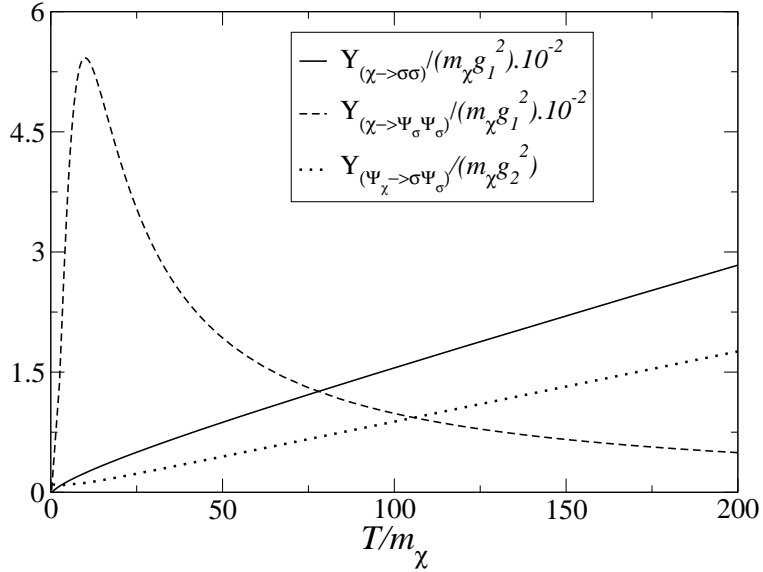}
\caption[]{\label{diss-ones} Individual contributions to the 
dissipation coefficient as a
  function of the temperature, for $g_1=g_2$, $h=h_1=h_2=h_3=0.5$, $g_3=0.01$,
  $m_\sigma=0.001 m_\chi$ and $m_{\psi_\chi}=m_\chi$,
  $m_{\psi_\sigma}=m_\sigma$.}
\end{figure}

It is also worth showing the individual contributions to the dissipation 
coming from the different intermediate catalyst fields
and the different decay channels 
available to them.
This is shown in {}Fig. \ref{diss-ones}.

Note from {}Fig. \ref{diss-ones} that for the
cases of interaction of a scalar intermediate field  with light
scalars and interaction of an intermediate  fermion with light scalar
and fermion, the high temperature behavior is linear in temperature.
Also, the  dissipation from the interaction of an intermediate scalar
field with light fermions dominates  in the regime of temperature
where $m_\chi \ll T \lesssim m_\chi/h_2$,  but for $T \gg m_\chi/h_2$,
the interaction of a intermediate scalar with light scalars dominates.
{}For  $T \gg m_\chi/h_2$ the contribution to dissipation  coming from
the interaction of an intermediate scalar field with light fermions
tend quickly to become subdominant with respect to both the interaction
of the intermediate heavy scalar $\chi$ with the light scalars
$\sigma$ and the intermediate heavy fermion with the light scalars and
fermions. 

Approximate semi-analytical fitting formulas for the dissipation, 
valid in both high-temperature regimes, with respect to
the masses of  the intermediate fields, $m \ll T \lesssim  m/h$ and $T
\gg m/h$, and expressed in terms of the  thermal masses for the
intermediate scalar and fermion fields, can be obtained by considering
separately the different  interaction forms.  Again, we cautioned that these
results at high temperature are only meant to be interpreted 
phenomenologically, sincethey are only derived at leading one-loop order.
The different cases are
listed below (for simplicity all expressions below were evaluated with
$m_{\chi_1} = m_{\chi_2}$):

a) {}Interaction of a heavy scalar into light scalar bosons, 
Eq. (\ref{Lchisigma}):

\begin{eqnarray}
&&\Upsilon \approx  0.21\frac{g_1^2}{h_1^2} T 
\left\{ 1 + 3 \left[ \frac{m_\chi(T)}{T}
\right]^{1/4} \right\}\;,
\label{UpsilonhighTB} 
\end{eqnarray}
where the thermal mass $m_\chi(T)$ is here defined as
$m_\chi^2(T) \simeq m_\chi^2 + g_3^2 T^2/12$, due to the interactions 
shown in Eq. (\ref{Lchisigma}).

b) {}Interaction  of a heavy scalar into light fermions, 
Eq. (\ref{Lchipsisigma}):

\begin{eqnarray}
&&\Upsilon \approx  7.3 \frac{g_1^2}{h_2^2 m_\chi} T^2 
\left[\frac{m_\chi}{m_\chi(T)} \right]^{3} \;,
\label{UpsilonhighTF} 
\end{eqnarray}
where the thermal mass $m_\chi(T)$ is here defined as
$m_\chi^2(T) \simeq m_\chi^2 + h_2^2 T^2/12$, due to the interactions shown 
in Eq. (\ref{Lchipsisigma}).

c) {}Interaction  of a heavy fermion into a light scalar boson 
and light fermion,
Eq.~(\ref{Lchipsichipsisigma}):

\begin{eqnarray}
&&\Upsilon \approx
5.53\times 10^{-5}\frac{g_2^2 T}{h_3^2} 
\left\{ 1 + 985.6 \left[ \frac{m_{\psi_\chi}(T)}{T}
\right]^{14/9} \right\}
\;,
\label{UpsilonhighTBF} 
\end{eqnarray}
where the thermal mass $m_{\psi_\chi}(T)$ is here defined as
$m_{\psi_\chi}^2(T) \simeq m_{\psi_\chi}^2 + h_3^2 T^2/12$, due to the 
interactions shown in Eq.~(\ref{Lchipsichipsisigma}).

As noted above from the results of {}Fig.  \ref{diss-ones}, the case
of interaction of the intermediate scalar field $\chi$ with boson
radiation fields $\sigma$ leads to growth linear in temperature of the
dissipation coefficient  at high temperature.  This differs from the
case of the scalar $\chi$ self-interacting, which found $1/T$ behavior
\cite{hosoya1,GR,BGR}.  The reason is, for the self-interacting case,
the decay width at lowest order was at two-loops, whereas for the
above case involving a separate radiation field, the decay width at
lowest order is at one-loop.  At one-loop the decay width is
proportional to an imaginary part of the self-energy that is ${\cal O}
(h_1^2 m_\chi^2)$, while at too-loop order, there are contributions
\cite{GR,BGR} of ${\cal O} (g_3^4T^2)$,  coming from the interaction
$g_3^2 \chi^2 \sigma^2$ and ${\cal O} (\lambda_\chi^2T^2)$, coming
from the $\chi$ field self-interaction, $\lambda_\chi \chi^4$.  This
difference in the decay rates at one-loop order and two-loops, results
in the differing temperature behaviors observed for the dissipation.
This difference in the leading behavior of the decay rates at one-loop
and two-loop orders also imply that the one-loop contributions are
dominant provided the temperature is not too high, that is, $T < {\rm
  min} (h_1 m_\chi/g_3^2, h_1 m_\chi/\lambda_\chi)$, which gives an
additional constraint for the consistency of the  derivations done
here based on the contributions of one-loop polarizations only. 

A crosscheck that the above estimates are at least parametrically correct even
in the absence of a full resummation treatment can be made by noting 
the close relation between the general dissipation coefficient expression,
Eqs. (\ref{cf}), (\ref{Sigmacorr0}), with those of the viscosity coefficients
derived through the Kubo formulas \cite{jeon,jeon1}. One observes that
aside for differences in powers of momentum in the defining integrals, the
expressions are similar. This similarity can be used
to relate the dissipation
coefficient with the known results for the viscosities, whose results,
found for example
in Refs. \cite{jeon,jeon1}, account for the full resummation of ladder
diagrams, just like the ones also contributing to the 
dissipation coefficient at high 
temperatures. This type of approach was used for instance by the author in 
Ref.~\cite{Bodeker:2006ij} to relate the moduli decay rate 
to the bulk viscosity
coefficient. This was possible because of the interaction form
used, which involved a derivative.  This allowed turning the calculation 
of the dissipation coefficient
in that reference to the analogous computation of the bulk viscosity 
through use of the Kubo formula.
In our case, following the lead of Ref.~\cite{Bodeker:2006ij},
the proper analogy would be instead to relate our expressions for the dissipation
with that for the shear
viscosity (and properly accounting for the difference in powers of momentum in the
integrals). This analogy was in fact raised before in Ref.~\cite{BGR}.
{}In the simplest case of the
interactions of the catalyst scalar 
field $\chi$ with only the radiation scalar field $\sigma$, 
this allow us to obtain an estimate for the
shear viscosity as $\zeta_{\rm shear} \sim T^5/(h_1^2 m_\chi^2)$. 
The different
temperature dependence with respect to the standard 
scalar field result obtained in the
$\lambda \phi^4$ theory~\cite{jeon}, $\zeta_{\rm shear} \sim T^3/\lambda^2$, 
comes from the different contribution to the decay width 
$\Gamma$ in each case
(as we noted in the previous paragraph, 
in the $\lambda \phi^4$ model $\Gamma$ comes from the imaginary 
part of the two-loop 
self-energy term, which is $ \sim \lambda^2 T^2$, while for the 
scalar catalyst-radiation
interaction we consider here, it is dominated by the one-loop 
self-energy computed
in the Appendix B). Relating the shear viscosity expression 
with our expression for the 
dissipation coefficient in this case (and accounting for the differences 
in factors of 
coupling constants and powers of temperature), in the high temperature 
regime we obtain 
$\Upsilon \sim g_1^2 T/h_1^2$. This is parametrically the result we found in 
Eq.~(\ref{UpsilonhighTB}).
Similarly, we could also relate the dissipation coefficient 
obtained for the other 
interaction forms. But this is still only meant to provide a parametric
estimate for the dissipation coefficients, since they are not the 
complete
resummation of all contributing higher order terms. Thus, we refrain ourselves
from following any further this similarity between dissipation and viscosity coefficients
in our case here, because it would also require a full calculation of the viscosity coefficients
as well, which is beyond the scope of the present paper.

\subsubsection{Contrasting the derivation of the high versus
low temperature dissipation coefficients}

At this stage it is useful to contrast the derivation of the dissipation
coefficients at high temperatures from this subsection
with those obtained in the 
low temperature regime of subsection \ref{lowTsubsec}.
One recalls from the calculations
done in the previous subsection \ref{lowTsubsec}, that 
in the low-temperature case, where what we called
the low-momentum approximation is valid, the dissipation
coefficients are proportional to the square of the decay widths.
This is evident from the behavior of the spectral  functions, such as in
Eqs. (\ref{rhochilowT2}), (\ref{rhopsichilowT2}) and from the explicit
expression for the dissipation 
coefficient in that regime,  Eq. (\ref{UpsilonlowT2}).
This very different behavior of the dissipation coefficients in 
the low-temperature regime is why they do not
suffer from the problems seen for the high temperature regime.
Due to this fact, it also means
no resummation is required in the
low-temperature case. 
{}Figure \ref{diss-T} provided one numerical example of parameter and
low-temperature regimes where this low-momentum approximation is
realized.    
As no resummation is needed,
the one-loop expression for the dissipation coefficients
in this regime alone provides reliable results. 

It is also instructive and easy to understand 
the low-temperature dissipation coefficients by looking at the
diagrams shown in {}Fig. \ref{multiloop}. 
The dissipation coefficients always carry catalyst propagators
in their upmost extreme points. Any higher order contribution to 
the dissipation
coefficient would then have at least four external 
heavy catalyst field propagators.
In this regime, the catalyst fields are at low-temperature, but the
light radiation fields are at high temperatures. Thus, the rungs made
with the light fields will be the ones that will have the most infrared
sensitivity and that can produce ${\cal O}(1)$ contributions as seen from
the
analysis made following  {}Fig. \ref{multiloop}.
But accounting for the at least four external
heavy catalyst fields in these diagrams, from the highest order contribution from them,
we are led to contributions of order
${\cal O} (\Gamma_i^4) \times {\cal O}(1)$, which are still of subleading order compared to 
the leading one-loop results for the low-temperature regime given
in Table \ref{tab1}, which recall are ${\cal O} (\Gamma_i^2)$. 
In the other case, two of these propagators can be just {}Feynman propagators,
which will lead to an overall suppression factor ${\cal O}(T^4/m_\chi^4)$ 
at low temperatures.
Since the dissipation coefficients in the low-temperature regime  
have direct application to the warm inflation scenario, where 
they were shown to be the most useful coefficients in model 
building calculations  \cite{BasteroGil:2009ec,BasteroGil:2006vr}, 
our results therefore have immediate utility.

\section{Applications}
\label{applications}

The dissipation coefficients computed in the previous sections
have numerous applications in cosmology and particle physics.  All these
coefficients have been computed with the system near thermal equilibrium and
under the adiabatic approximation  Eqs. (\ref{adiabatica}) and
(\ref{adiabatict}) with the microscopic timescale $\tau$ in these
equations identified with the inverse decay widths $\Gamma_i$.  In the adiabatic
approximation all microscopic physics, that governs the creation of
dissipation, operates much faster than any macroscopic changes
that the dissipation affects.  In particular, 
the consistency of the adiabatic approximation
requires that the decay widths that enter the
dissipation coefficients must be much bigger than the motion of
in the  background scalar field and changes in temperature \cite{BGR},
\begin{equation}
\Gamma_i \gg \frac{\dot \phi}{\phi} , \frac{\dot T}{T} .
\label{adiab1}
\end{equation}
In the cosmological setting, in addition the adiabatic approximation requires
\begin{equation}
\Gamma_i \gg H ,
\label{adiab2}
\end{equation}
where $H$ is the Hubble parameters.  
When the conditions (\ref{adiab1}) and (\ref{adiab2}) apply, we can 
readily use the results for the dissipation
we have obtained. 
Below some systems are discussed, where
these conditions apply and dissipation effects are important.

\subsection{Thermalization phase ending reheating}

In the standard inflation picture \cite{ci},  during inflation the
universe supercools. In order to end inflation and enter the subsequent
radiation dominated regime, a reheating phase is required in which coupling of
the inflaton to other fields becomes important and coherent oscillations of
the inflaton induce particle production in these coupled fields
\cite{reheatu}. There has been exhaustive study of the phases of reheating.
In many models there is a first phase, preheating \cite{Kofman:1994rk}, where
there are large oscillations of the inflaton, which through parametric
resonance lead to explosive particle production.  Once the inflaton
oscillations decrease,  this leads to the second phase of small coherent
oscillation  leading to a perturbative reheating phase \cite{reheatu}.  In all
reheating processes, there is one final phase, where the inflaton oscillation
motion ends, the radiation that has been produced is thermalized, and then the
radiation dominated epoch begins.  This can be called the thermalization phase
of reheating.  
In a generic reheating situation the oscillations of $\phi$ are damped due to
the Hubble damping term, thus generally  ${\dot \phi}/\phi$ is a decreasing
function of time.  Thus at some point the adiabatic condition
Eq. (\ref{adiab1}) is realized for particle type $i$. And depending on  the
strength of the coupling, the other adiabatic conditions Eq. (\ref{adiab2})
also are realized.  This is the adiabatic regime where the macroscopic 
motion of
the inflaton field and expansion of the universe becomes slow relative to the
microscopic motion.  Once that occurs, and provided  $\phi$ is coupled to
other fields, its evolution equation has the form
\begin{equation}
{\ddot \phi} + (3H + \Upsilon) {\dot \phi} + \frac{d V}{d \phi} = 0,
\label{pddote}
\end{equation}
where depending on how $\phi$ is coupled to the other fields, $\Upsilon$ are
some combination of the dissipative coefficients computed in the earlier
sections.  Here the dissipative term will lead to particle production, with
the radiation energy density governed by
\begin{equation}
{\dot \rho}_r + 4 H {\rho}_r = \Upsilon {\dot \phi}^2 .
\label{rdote}
\end{equation}
Thus, this equation will control the final temperature of the universe after
reheating. 

The dissipative coefficient already requires  the presence of a  thermal bath
during reheating. In general this is not difficult to achieve, and preheating
for example can serve to that end. Being a non-perturbative process, even
particles heavier than  the inflaton may be produced, which then decay into
lighter degrees of freedom. After these particles thermalize, the initial
thermal bath is produced.  However preheating does not always occur and even
when it does, typically it is not enough to convert all the vacuum energy from
inflation into radiation.  The next phase is standard reheating, in which the
perturbative decay of the inflaton occurs.  This phase is generic and produces
an initial thermal bath.  Eventually as the inflaton motion slows and the
thermal bath grows the final thermalization phase starts, in which particle
production occurs through dissipation governed by Eqs. (\ref{pddote}) and
(\ref{rdote}).  

That the adiabatic condition Eq.(\ref{adiab1}) is reached at some point during
the oscillations is practically ensured in most of the models of
inflation. {}For example, for the chaotic inflation model $\lambda \phi^4$,
during the oscillations it is found that $\phi$ slows as $\phi \sim a^{-1}$
and ${\dot \phi} \sim a^{-2}$, where $a(t)$ is the cosmic scale factor
\cite{gkls}.  Thus ${\dot \phi}/{\phi} \sim a^{-1}$ is decreasing,  and 
provided that at least $\Gamma_i\approx {\rm const.}$ or growing, 
which is often the case, then
eventually the  adiabatic condition will
be satisfied. On the other hand if the energy density of the oscillating
inflaton behaves as  matter, like in quadratic chaotic models or in hybrid
inflation, then ${\dot \phi}/{\phi} \sim {\rm const.}$. Nevertheless,
generically one has $\dot \phi / \phi \sim m_\phi$, $m_\phi$ being the
inflaton mass and the typical value for the frequency of the oscillations, and
the condition to start in the adiabatic regime reads $\Gamma_i > m_\phi$. This
is not difficult to fulfill for the decay rate of a heavier field than the
inflaton, with for example $\Gamma_i \simeq h_2^2m_i/(8 \pi)$ being the
standard perturbative decay into massless fermions for a particle type $i$
with mass $m_i \gg m_\phi$.  

In other inflationary scenarios, the inflaton does not undergo a series of
damped oscillations about a potential minima, but inflation is followed
instead by a kination period \cite{kination}, and reheating takes place
through gravitational particle production \cite{ford}.  If kination lasts long
enough, at some point radiation takes over  the kinetic energy density of the
inflaton and the universe will enter the radiation dominated epoch with a
temperature $T_{RH} \simeq 10^3 N_s^{3/4}\, GeV$, where $N_s$ is the number of
scalar degrees of freedom gravitationally produced. However, the same
gravitational mechanism also produces a stochastic background of gravitational
waves, and the overproduction of gravity waves is one of the potential
problems of this kind of scenarios. The Big Bang Nucleosynthesis (BBN) bound
on the energy density of these gravity waves translates then into a lower bound
for $N_s > O(100)$. On the other hand, during kination the inflaton field only
evolves logarithmically with the scale factor, and then ${\dot  \phi}/{\phi}
\sim a^{-3}$. That is, the adiabatic regime can be quickly achieved. Kination
can end before the gravitationally produced radiation takes over, and it will
be followed instead by radiation production through dissipation. The shorter
kination period means   less gravity waves, that could be kept within BBN
bounds without any further constraint on the parameters. 

\subsection{Warm inflation}

The standard picture of inflation is where the universe supercools
during the inflation phase, since there is no particle production.
Such a cold inflation picture assumes the inflaton has negligible
coupling to other fields during the inflation phase.  However in a
realistic particle physics model, the inflaton field could be coupled
to other fields. In this case an alternative dynamical realization of
inflation is possible, which is warm inflation, where particle
production occurs concurrent to inflationary expansion
\cite{Berera:1995ie,wi1,wi2}, with particle production occurring due to
the interaction of the inflaton with other fields.  The presence of
particles then leads to a sustained heat bath.  This heat bath
influences the dynamics of dissipation as well as acts on the inflaton
fluctuations.  The statistical state of the heat bath could be any
nonequilibrium state, and a field theoretic dynamical calculation is
necessary.  However in treatments up to now, an assumption of
thermalization is used, which is then checked self-consistently in the
calculation.  In such a case, the fluctuations of the inflaton, which
are the primordial seeds of density perturbations, are thermal in
origin \cite{im,Berera:1995wh,Berera:1999ws}.  Several models of warm
inflation under this picture have been studied
\cite{Berera:1999ws,BR1,BMR,BR21,BR22,BasteroGil:2006vr,BasteroGil:2009ec,wimodels,pavon}.

Since inflationary expansion requires the $\phi$ motion to be slow, the
adiabatic approximation Eq. (\ref{adiab1}) can be applicable depending on the
coupling of $\phi$ to other fields.  In such a regime, and also provided the
second adiabatic condition Eq. (\ref{adiab2}) is satisfied, the evolution
equation for the inflaton is governed by Eq. (\ref{pddote}), which in the
slow-roll  regime becomes
\begin{equation}
(3H + \Upsilon) {\dot \phi} + \frac{dV}{d \phi} \simeq 0 ,
\label{pdote}
\end{equation}
with particle production governed by Eq. (\ref{rdote}).  Under these
conditions, along with the General Relativity condition $\rho_v > \rho_r$,
where $\rho_v$ is the vacuum energy density, a warm inflationary expansion
occurs.  Moreover typically $m_{\chi}$, $m_{\psi_\chi} \gg H$, in which case
the dissipative coefficients computed in the earlier sections can be applied
to cosmology.  Under these conditions and depending how $\phi$ couples to the
other fields in the underlying particle physics model, $\Upsilon$ in the above
equation represents some combination of dissipative coefficients computed in
the previous sections.

The most successful models of warm inflation~\cite{BMR}
use the two-stage mechanism \cite{BR1}, in which the inflaton field
interacts with heavy intermediate fields, e.g. like through the interaction
terms (\ref{Lphichi}) and (\ref{Lphipsichi}).  These intermediate heavy
fields, which can be either bosons or fermions, in turn interact with light
fields, which can also be either bosons or fermions, with interaction terms
realizing the two-stage mechanism \cite{BR1}
such as Eqs. (\ref{Lchisigma}), (\ref{Lchipsisigma}) or
(\ref{Lchipsichipsisigma}).  At leading one-loop order, these decay rates can
be expressed e.g. in terms of Eqs. (\ref{Gammachi}) and (\ref{ImSigmapsi}),
for the interaction terms  (\ref{Lchisigma}), (\ref{Lchipsisigma}) and
(\ref{Lchipsichipsisigma}).

Note also that since in
warm inflation the heavy intermediate fields have masses $m_\chi, m_{\psi_\chi}
\gg H$ and even the light fields, $\sigma$, $\psi_{\sigma}$,
will generally have masses
larger than $H$,  
the effects of the expansion can be neglected in the expressions
involving the respective field propagators and, 
in particular, the expressions for the dissipation 
coefficients can be well approximated by their Minkowski
expressions.   

Once the distinguishing feature of the dynamics in warm inflation is 
due to dissipation, knowing precisely the dissipation coefficients, their
temperature dependence and their regime of validity is fundamental.
This was recently exemplified in the Refs. \cite{mossgraham,shearpert},
where it was shown how the temperature dependence of the dissipation coefficient
can affect the power spectrum of perturbations. 
The results for the dissipation coefficients and the derivation
of their explicit temperature dependence as we have obtained  here becomes then
of fundamental relevance in the precise determination of the
power spectrum in warm inflation. 

Dissipative effects might also be relevant immediately after warm
inflation ends, during the first stages of the reheating 
process, when different production mechanisms are at work. 
For example a network of cosmic strings is formed at the end of the
so-called D-term SUSY hybrid inflation \cite{cosmicstrings}, 
in which a $U(1)$ symmetry is spontaneously broken.
If cosmic strings form after a period
of warm inflation, the transition will take place with the fields
place in a thermal bath, and/or with dissipation still
operative. This may affect the rate of formation of the cosmic
strings network, and their amplitude contribution to the
CMB. Observations limit their contribution to be less than a 10 \% of
the total amplitude \cite{cosmiclimits}, which translates as
usual into constraints on the model parameters.  Typically the 
constraint comes from having an inflation scale that is
too large, which
translates into too large of an amplitude of the primordial spectrum
generated by the cosmic strings. The constraint can therefore be relaxed by
a change of the potential which leads to a decrease of the inflationary
scale. For 
example by considering a SUSY model with a non-minimal K{\"a}hler
potential as in Ref. \cite{seto}. Similarly, given that warm inflation typically
requires a lower inflationary scale, it may also help in reducing the
contribution to the spectrum of the cosmic strings. 

 If the symmetry broken
 during the phase transition is instead a discrete symmetry, this will lead
 to the formation of a network of Domain Walls (DW). When the DW
 are stable,  these pose a severe cosmological problem, as the energy
 of the domain wall network will tend to dominate the total energy
 density, and they induce large $T$ fluctuations in the CMB
 \cite{DWproblem}. These can 
 be avoided if at the time of the transition one of the vacua is
 favored over the other, for example by having a tilt in the
 potential, or a biased initial configuration, towards one of the
 vacua \cite{Gelmini:1988sf} such the DW evolve and annihilate after
 formation. Other solution can be to dilute them by having a extra
 period of  inflation  after the phase  transition;  of the order of
 $O(10)$ e-folds of rapid  expansion would be enough to dilute whatever
 unwanted relics have been produced 
 after a normal period of inflation. This can be achieved for example
 with thermal inflation \cite{thermal}, where inflation is driven
 by the $T$ corrections restoring the symmetry and keeping the field
 in the false vacuum at a constant  vacuum energy. Similarly, given
 the appropriate  interactions in the model, once the adiabatic conditions are
 fulfilled for having inflation driven by dissipation, it may be
 enough to have some e-folds of warm inflation to dilute the network
 of DW. There will be less restrictions on the
 parameters values given that one does not require a full period of
 inflation and does not demand to comply with the CMB constraint on
 the primordial spectrum. Notice however that while thermal
 inflation is driven by the $T$ corrections due to direct couplings
 of the inflaton to light fields, in warm inflation
 the extra bout of inflation will be driven by dissipative effects in
 a thermal bath, while the field is moving towards the true minimum
 of the potential, but with the thermal corrections to the inflaton potential
 being negligible. 

\subsection{Second order phase transitions}

Several relativistic quantum field theory systems exhibit second order phase
transitions.  U(1) model of scalar field coupled to gauge field is a common
example \cite{Martin:1994zg}.   The electroweak phase transition in certain
regimes, when $m_{Higgs} \stackrel{>}{\sim} m_{gauge}$ has been argued
exhibits a second order transition for many  models
\cite{MarchRussell:1992ei}.  The chiral phase transition is also believed to
be second order in QCD for two flavors of massless quarks
\cite{Rajagopal:1992qz}.  Several other examples of second order phase
transitions in quantum field theory can also be found \cite{sop}.

The study of second order phase transitions often  focus on the
phenomenological Landau-Ginzburg mean field theory such as in
\cite{MarchRussell:1992ei} or utilize a Langevin equation as in
\cite{Martin:1994zg}.  Most of these treatments rely on symmetry arguments and
general principles to analyze the transition, but fall short of a first
principles calculation. For that the full treatment of all interactions is
required, which would then generate dissipative effects of the
scalar order parameter.  For that, the coefficients computed in this paper
are applicable.

\section{Summary and Conclusions}
\label{summary}

In this paper dissipative  coefficients have been
computed in the near thermal equilibrium and adiabatic approximation,
Eqs. (\ref{adiab1}), for all combinations of interactions between
scalar and fermion fields.  There are several new results here.  In
particular, several new interactions presented in
Sect. \ref{interactions} have been considered, which have not
previously been treated  in the literature for these types of
calculations.  {}For the dissipation coefficients at low temperature
this has led to sub-leading terms in temperature in
Eq. (\ref{UpsilonlowT}).    At high
temperature, the interactions structures we have studied
lead to linearly growing dissipation
coefficient, which differs from the $1/T$ behavior found in earlier
papers \cite{hosoya1,GR,BGR}, where only scalar field self-interaction
and interaction with another light boson field were considered.

At high temperature, as is well known and we have noted, 
there are resummation issues with
the derivation of the dissipation coefficients at leading one-loop order.
In particular there are higher order loop 
terms giving contributions that are of the same order 
as the one-loop term. We have discussed that these higher loop terms 
can in principle be treated in the same way as in the derivation of viscosity
coefficients, which also show the same need for resummations and have
been treated in the literature.
However, due to the complicated nature of the multifield 
interactions we have used 
in this paper, this is expected to be a much more difficult task.
Even so, we have provided estimates for the dissipation coefficients in the
high temperature regime, which should be interpreted only phenomenologically,
giving an order of magnitude estimate for them.

On the other hand, in the low-temperature regime, where the catalyst
fields have masses much bigger than the temperature scale 
while the radiation fields are at
high temperatures, we have shown that the dissipation coefficients
in that regime do not suffer from the same resummation issues and the results
we have provided in this regime are precise. 
This is a direct consequence of  multi-field type
interactions treated here. The freedom in coupling and mass parameters avails
itself so that in certain parameter regimes, calculations
only up to one-loop order of
the coefficients are adequate, without requiring resummation of higher
order terms, as usually required in the evaluation of these
coefficients in single scalar field self-interaction or gauge field cases.
In particular by working with multiple types of interactions involving
intermediate fields and light fields, two independent coupling
parameters occur, thus allowing for regimes where higher order
contributions are sub-leading. This is a  huge advantage in the use of
these type of interactions  as compared for example with computing
transport coefficients in the single field  case.

There are several applications of these coefficients, and the interaction
structures we have studied in cosmology, as explained in the last section,
but other applications can also be found in the context of
heavy ion collision, relativistic phase
transition and any other case that necessarily involves dissipative
processes related to a background or order parameter field used to study
the dynamics. Many treatments using these coefficients can be found in
the literature for study of warm inflation, chiral condensate in heavy
ion collision, and study of phase transitions.  In
Sect. \ref{applications} we have identified a few more  new
applications.  In particular, we have observed that the final stages
of reheating after inflation generically will lead  to an adiabatic
evolution of the inflaton field.  During this period, which we have
termed the thermalization phase after reheating, the dissipation
coefficients computed here and related past papers are important, and
will assist the inflaton in ending its oscillations, exiting entirely
from reheating, and finally leaving the  universe in a thermal
equilibrium state.   In a different direction, for inflation scenarios
ending with a kination period with reheating through gravitational
particle production, we have identified that dissipation effects could
be important, thus decreasing the length of this period, which can
have consequences on gravity wave production.  

We have computed
the contributions to dissipation from various types of interaction,
involving fields with very different mass scales.
{}First, since the sorts
of interactions we consider in this paper are common in many models,
the contributions from them to dissipation would
still need to be computed to understand how significant they are, and
this has been achieved here. 
Moreover, there are very few exact limits or approximations
in which dissipation coefficients can be computed.
Table \ref{tab1} contains results from this new low-momentum
approximation, first developed in \cite{mossxiong0}.
As such, the results in Table \ref{tab1}
might prove useful as checks for numerical
algorithms for computing  dissipation coefficients.

\appendix

\section{Model building}
\label{models}

In this Appendix we give some model examples for the interactions
considered in section \ref{interactions}. We have focused on a
scenario with several scalar fields and fermions; one of
the scalars, $\phi$, 
acquires a background value $\varphi(t)$,  and couples to a set of boson
$\chi$ and fermion $\psi_\chi$ fields. The latter acquire
masses through their interaction with $\phi$. The $\chi$ 
and $\psi_{\chi}$ fields in turn
couple to another set of boson $\sigma$ and fermion $\psi_\sigma$
fields, which are kept as 
light fields. By definition therefore this set of light fields cannot
couple to $\phi$, otherwise they will pick up a mass similarly to 
$\chi$ and $\psi_\chi$. The coupling between the
scalar field with non-vanishing background value and a set of light fermions
can be easily forbidden by imposing an U(1) global symmetry. On the
other hand, the coupling between the $\phi$ field and the light
scalars can be avoided in a supersymmetric model plus a
global or discrete symmetry. Supersymmetry protects the light sector
from getting large mass corrections.  

\subsection{Supersymmetric model}

A supersymmetric model with the pattern of interactions considered in
section \ref{interactions} is given for example by the superpotential:
\be
W= g \Phi X^2 + h X Y^2 \,,
\ee
where $\Phi$, $X$, $Y$ are superfields with scalar and fermion
components given by ($\phi$, $\psi_\phi$), ($\chi$, $\psi_\chi$) and
($\sigma$, $\psi_\sigma$) respectively. The scalar potential is given
by:
\be
V= g^2 |\chi|^4 + h^2 |\sigma|^4  
+ 4 g^2 |\chi|^2 |\phi|^2 + 4 gh Re[\phi^\dagger \chi^\dagger \sigma^2] + 4 h^2
|\chi|^2 |\sigma|^2 \,,
\ee
whereas the Yukawa interactions are given by:
\be
-{\cal L}_F= 2 g \phi \bar \psi_\chi P_L \psi_\chi  + g \chi \bar
\psi_\phi P_L \psi_\chi + 2 h\chi \bar \psi_\sigma P_L \psi_\sigma
+  h\sigma \bar \psi_\chi P_L \psi_\sigma + h. c. \,.
\ee
Comparing with the interactions given in section \ref{interactions}
we have for the couplings:
\bea
g_1&=& g_2 = 2 g \,,\\
g_3 &=&  2 h_1 =  h_2 = 2 h_3 = 2 h \,, \\
\lambda_\chi &=& g^2 \,,\;\;\; \lambda_\sigma=h^2\,, \;\;\; \lambda_\phi=0
\,, 
\label{susycouplings}
\eea
and once $\phi$ gets a background value $\varphi/\sqrt{2}$, the mass
parameters are given by:
\be
M= \sqrt{2} g \varphi\,, \; \;\; m_{\chi_1}= m_{\chi_2}= m_{\psi_\chi}=
\sqrt{2} g \varphi \,,     
\ee
with $m_\sigma=m_\phi=m_{\psi_\phi}=m_{\psi_\sigma}=0$. We recover the
interaction terms given in section \ref{interactions}, 
plus one additional interaction between the boson $\chi$, one heavy
fermion $\psi_\chi$ and one light fermion $\psi_\phi$. Due to the
presence of the massive fermion in this term, its contribution to the
dissipation  coefficients will be subdominant
(suppressed by extra powers of $m_\chi$) with respect to that given by
the interaction of $\chi$ with two massless fermions $\psi_\sigma$. 

The conditions on the couplings given in Eq. (\ref{conditions1}) are
fulfilled by choosing $h \ll 1/2$ and $g^2 \ll h$. On the other hand,
the quartic coupling $\lambda_\sigma$ is the same than $g_3^2=  
h_2^2 = 4 h_3^2$. Therefore, in the high-$T$ regime, the temperature dependent
corrections to the masses of $\sigma$ and $\chi$ are comparable.  
On the other hand, in the low-$T$ regime with $m_\chi/T
>O(10)$, keeping the $\sigma$ field light when including thermal
corrections due to its self-coupling only requires $\lambda_\sigma =
h^2 \ll m_\chi^2/T^2$,  which is fulfilled with $h \ll 1/2$. 

So far we have assumed that SUSY is unbroken except by thermal
corrections; i.e, from the Lagrangian we have $M= m_\chi=
m_{\psi_\chi}$, and $m_\sigma=0$. However, thermal corrections will
split the boson-fermion masses for the $X$ superfield, with
\be
\Delta m_\chi^2= |m_{R,\chi}^2 - m_{R,\psi_\chi}^2| = \frac{g^2 + h^2}{3} T^2 \,,
\ee
and this is in turn can induce a mass term for the light field
$\sigma$ at one loop,  
\be
\delta m_\sigma^2 \sim \frac{h^2 \varphi^2}{(4 \pi)^2} \left(
\frac{\Delta m_\chi^2}{m_\chi^2} \right)\,.
\ee  
Therefore, an additional condition on the couplings has to be imposed for 
the $\sigma$ field to remain light, $\delta m_\sigma^2 \ll T^2$,
in the low-$T$ regime: 
\be
\frac{h^2}{ 3 (4 \pi)^2} \left( 1 + \frac{h^2}{g^2} \right) \ll 1 \,.
\ee
Taking into account that we have already required $h^2 \gg g$, this reduces
the allowed range for the couplings, but still with a mild hierarchy of the
order of $h^2/g \sim O(10)$  all the constraints are fulfilled. 
On the other hand, SUSY can be already broken independently of thermal
corrections, and in this case we also need to impose $\Delta
m_\chi^2/T^2 \ll (4 \pi)^2 g^2/h^2$. In addition, SUSY breaking may
induced additional trilinear couplings in the scalar potential such
that now $M> m_\chi$.

To summarize so far, in a SUSY model where the interactions between
heavy and light fermions are given in terms of two coupling $g$ and
$h$, the consistency conditions for the expressions computed in the
low-$T$ regime given in Table \ref{tab1} to be valid are:
\be
h \ll 1/2 \,\;\;\;, \frac{h^2}{g} \simeq O(10) \,.
\ee  

\subsection{Non-supersymmetric model}

In a non-supersymmetric model, scalar masses are not protected, and
with the couplings $|\chi|^2|\phi|^2$ and $|\chi|^2|\sigma|^2$ in the
potential, a mass term for $\sigma$ may be radiatively generated through

an effective vertex $|\sigma|^2 |\phi|^2$. Thus, although
symmetry breaking patterns can be chosen such that a sector of the
boson fields remains massless (the Goldstone bosons), those will
couple to the $\phi$ field. 

Instead, the scenario worked out in this paper will correspond to a model
with two coupled scalar fields, $\phi$ and $\chi$, and two sets of
fermions: $\psi_\chi$ couples to $\phi$ and it is massive, whereas
$\psi_\sigma$ only couples to $\chi$. Yukawa interactions are
therefore:
\be
-{\cal L}_F = g_2 \phi \bar \psi_\chi P_L \psi_\chi  + h_2 \chi \bar
\psi_\sigma P_L \psi_\sigma + h. c. \,,
\ee
and the scalar potential is given by:
\be
V= \lambda_\phi |\phi|^4 + \lambda_\chi |\chi|^4 
+g_1^2 |\chi|^2 |\phi|^2 + \mu_\phi^2 |\phi|^2 + \mu_\chi^2 |\chi|^2 \,.
\ee
Taking  the mass parameters $\mu_\phi^2<0$ and $\mu_\chi^2>0$, the
field $\phi$ acquires a vacuum expectation value $\varphi/\sqrt{2}$,
and the scalar fields  get  field dependent masses $m_\chi^2= \mu_\chi^2 +
g_1^2 \varphi^2$ and $m_\phi^2= \mu_\phi^2 + 3 \lambda_\phi
\varphi^2$. In addition, we impose $m_\phi \ll m_\chi$, i.e.,
$\lambda_\phi \ll g_1^2$. Therefore, the $\phi$ field is a light field
with a small self-interaction. In the absence of any other light field, dissipation 
is then given by the fermionic
contribution. In the low-$T$ regime this is the second line in Table
\ref{tab1}; and in the high-$T$ regime, it is given by
Eq. (\ref{UpsilonhighTF}).

\section{Decay widths}
\label{app}

Below we give some of the details of the calculation of the relevant
decay widths used
in the expressions for the dissipation coefficients.  The
particle decay widths can be expressed as usual in terms  of the imaginary
parts of the self-energies diagrams~\cite{bellac}.  The relevant self-energy
terms in our calculations come from  one-loop processes and these can all be
analytically derived in detail.  There are three main processes we consider:
(1) decay of a scalar boson to scalar bosons; (2) decay of a
scalar boson to fermions and (3) decay of a fermion  to a 
scalar boson and a fermion.

\subsection{decay of a
scalar boson to scalar bosons}

 Let us first consider the decay of a scalar boson $X$ to two other scalar
 bosons, e.g. $X \to X_1+X_2$, with an interaction vertex given
 by $g_X X X_1 X_2$. The imaginary part
 of the self-energy for the leading order contribution is given by (see also
 \cite{weldon})

\begin{equation}
{\rm Im} \Sigma_\chi(p_0, {\bf p}) = \frac{g_X^2}{2} 
\left(e^{\beta p_0} -1\right) \int \frac{d^4 k}{(2 \pi)^4} 
n_B(p_0-k_0) \rho_{X_1}(p_0-k_0, {\bf p}-{\bf
  k})n_B(k_0)
\rho_{X_2}(k_0, {\bf k})  \;,
\label{ImSigmachi}
\end{equation}

\noindent
where $n_B$ is the Bose-Einstein distribution function
and $\rho_{X_i}(k_0, {\bf k})$ is the free spectral  function for the
$X_i$ (i=1,2) scalar field,

\begin{equation}
\rho_{X_i}(k_0, {\bf k}) = 2 \pi \delta(-k_0^2 +  \omega_{X_i}^2({\bf
  k}))\;,
\label{specsigma}
\end{equation}

\noindent
with $\omega_{X_i}({\bf k}) = \sqrt{{\bf k}^2 + m_{X_i}^2}$.  The
Dirac-delta functions in Eq. (\ref{ImSigmachi}) give the different implicit
absorption and decay processes involving the $X,X_1$ and $X_2$ particles,
including the different branch cuts. Taking for example $p_0>0$ (for $p_0 <0$
there is a change of sign in the expressions) and also assuming $m_{X_1} \geq
m_{X_2}$,  there are two processes
contributing: one satisfying $p_0 = \omega_{X_1}({\bf p}-{\bf k})
+\omega_{X_2}({\bf k})$, for $p_0^2 - {\bf p}^2 > (m_{X_1}+m_{X_2})^2$,
corresponding to decay of an off-shell $X$ particle into on-shell 
$X_1$ and $X_2$
particles, and the other for $p_0 + \omega_{X_2}({\bf k}) =
\omega_{X_1}({\bf p}-{\bf k})$,  for $ - |m_{X_1}^2-m_{X_2}^2| 
\leq p_0^2 - {\bf p}^2  \leq (m_{X_1}-m_{X_2})^2$, corresponding to
Landau damping, where an off-shell particle scatters with two on-shell
particles from the heat bath. Explicitly, for the decay contribution

\begin{equation}
{\rm Im} \Sigma_\chi^{\rm decay}(p_0, {\bf p}) = g_X^2 \pi
\int \frac{d^3 k}{(2 \pi)^3}
\frac{1}{4 \omega_{X_1}({\bf p}-{\bf k})\omega_{X_2}({\bf k}) } \left[ 1 + 
  n_B(\omega_{X_1}({\bf p}-{\bf k})) + n_B(\omega_{X_2}({\bf k}))\right] 
\delta(p_0 - \omega_{X_1}({\bf p}-{\bf k})
-\omega_{X_2}({\bf k}) ) \;,
\label{Sigmachidecay}
\end{equation}
and for the Landau damping contributions,

\begin{equation}
{\rm Im} \Sigma_\chi^{\rm LD}(p_0, {\bf p}) = g_X^2 \pi  
\int \frac{d^3 k}{(2 \pi)^3}
\frac{1}{4 \omega_{X_1}({\bf p}-{\bf k})\omega_{X_2}({\bf k}) }
\left[ -n_B(\omega_{X_1}({\bf p}-{\bf k})) + n_B(\omega_{X_2}({\bf k}))\right] 
\delta(p_0 - \omega_{X_1}({\bf p}-{\bf k})
+\omega_{X_2}({\bf k}) ) \; .
\label{SigmachiLD}
\end{equation}

The angular integration in Eqs. (\ref{Sigmachidecay}) and (\ref{SigmachiLD})
can be done using the Dirac-delta functions. Using the formula,

\begin{equation}
\delta(f(x)) = \sum_i \frac{\delta(x-x_i)}{\bigl|\frac{df}{d {x}} \bigr|}
\;,\;\;\;f(x_i)=0\;,
\end{equation}
we obtain for example that

\begin{equation}
\delta(p_0 - \omega_\sigma({\bf k}) -
\omega_\sigma({\bf p}-{\bf k})) =
\frac{\omega_{X_1}({\bf p}-{\bf k})}{|{\bf p}||{\bf k}|} \delta( \cos\theta -
\cos\theta_0)\;,
\end{equation}
where $\cos \theta_0$ is the solution of 
$p_0 -  \sqrt{{\bf p}^2+{\bf k}^2 -2|{\bf p}||{\bf k}|\cos\theta_0
  + m_{X_1}^2}- \sqrt{{\bf k}^2 +
  m_{X_2}^2}  =0$. This restricts the limit of the momentum integration to
the values such that $-1 \leq \cos\theta_0 \leq 1$, or $k_- \leq k \leq k_+$,
where,

\begin{equation}
k_\pm^2 = \frac{1}{4}\left\{ |{\bf p}| \left( 1 +
\frac{m_{X_1}^2-m_{X_2}^2} {-p_0^2+{\bf p}^2} \right) \pm p_0 \left[
  \left( 1 + \frac{m_{X_1}^2-m_{X_2}^2} {-p_0^2+{\bf p}^2} \right)^2
  + \frac{4 m_{X_2}^2}{-p_0^2+{\bf p}^2}\right]^{1/2}  \right\}^2\;,
\end{equation}
and the integration over the momentum in
Eqs. (\ref{Sigmachidecay}) and (\ref{SigmachiLD}) is now trivial.
Defining $\omega_\pm$ by
\begin{equation}
\omega_\pm = \sqrt{k_\pm^2+m_{X_2}^2}\;,
\label{omegapm}
\end{equation}
and by the definition of
the decay width, Eq. (\ref{Gammachi}), 
the (one-loop) result coming from the $X \to X_1 + X_2$ process
involving scalar bosons is

\begin{eqnarray}
\Gamma_{X}(p_0,{\bf p}) &=& g_X^2
\frac{1}{32 \pi \omega_{X}({\bf p})}  \left\{ \frac{\omega_+ -
\omega_-}{|{\bf p}|}
+  \frac{T}{|{\bf p}|} {\rm ln} \left[ \left(
  \frac{ 1- e^{-\frac{\omega_+}{T}} }{ 1- e^{-\frac{\omega_-}{T}} } \right)
\left(
\frac{ 1- e^{-\frac{p_0-\omega_-}{T}} }{ 1- e^{-\frac{p_0-\omega_+}{T}}
} \right)
  \right] \right\} \theta(p_0^2 - {\bf p}^2 - (m_{X_1} + m_{X_2})^2) 
\nonumber \\ &+&
 g_X^2
\frac{1}{32 \pi \omega_{X}({\bf p})}  
\frac{T}{|{\bf p}|} 
{\rm ln} \left[ \left(
  \frac{ 1- e^{-\frac{\omega_+}{T}} }{ 1- e^{-\frac{\omega_-}{T}} } \right)
\left(
\frac{ 1- e^{-\frac{p_0+\omega_-}{T}} }{ 1- e^{-\frac{p_0+\omega_+}{T}}
} \right)
  \right]  \theta(-p_0^2 + {\bf p}^2 + (m_{X_1} - m_{X_2})^2) 
\;,
\label{Gamma1}
\end{eqnarray}
where $g_X^2$ has mass dimension two.

\subsection{Decay of a 
scalar boson to fermions}

The second process considered is the decay of a scalar boson $X$ to two
fermions. We follow for simplicity the same notation
as used in the previous case, but now $X_1$ and $X_2$ as a 
fermion-antifermion pair, which has a Yukawa coupling $g_X$ with the 
scalar boson $X$.
The imaginary part of
the one-loop $X$ field self-energy is now given by

\begin{equation}
{\rm Im} \Sigma_{\chi}(p_0, {\bf p}) = g_X^2   \left(1-e^{\beta p_0} \right)
\int \frac{d^4 k}{(2 \pi)^4} n_F(p_0-k_0)  n_F(k_0) {\rm
  tr}\left[ \rho_{X_1}(p_0-k_0, {\bf
    p}-{\bf k}) \rho_{X_2}(k_0, {\bf k})  \right]\;,
\label{ImSigchi}
\end{equation}

\noindent
where $n_F$ is the {}Fermi-Dirac distribution function and
$\rho_{\psi_\sigma}(k_0, {\bf k})$ is the free spectral  function for the
fermion field,

\begin{equation}
\rho_{X_i}(k_0, {\bf k}) = 2 \pi \left( -\not\!k +
m_{X_i}\right)\delta(-k_0^2 +  \omega_{X_i}^2({\bf k}))\;,
\label{specpsi}
\end{equation}
with $\omega_{X_i}({\bf k}) = \sqrt{{\bf k}^2 + m_{X_i}^2}$.

As in the previous decay mode case, when (\ref{specpsi}) is used in
(\ref{ImSigchi}), we can again identify decay and Landau damping terms.  The
momentum integration is carried out also in a very similar way, giving as the
final result, from the definition  Eq. (\ref{Gammachi}), 
that the decay width of a scalar boson $X$ decaying into  a
fermion-antifermion pair $X_1+X_2$ is given by

\begin{eqnarray}
\Gamma_{X}(p_0,{\bf p}) &=&  g_X^2
\frac{-p_0^2 + {\bf p}^2}{8 \pi \omega_{X}({\bf p})}  
\left[1 - \frac{(m_{X_1} + m_{X_2})^2}{p_0^2 - {\bf p}^2}\right]
\left\{ \frac{\omega_+ -
\omega_-}{|{\bf p}|} \right.
\nonumber \\
&+& \left.  \frac{T}{|{\bf p}|} {\rm ln} \left[ \left(
  \frac{ 1+ e^{-\frac{\omega_+}{T}} }{ 1+ e^{-\frac{\omega_-}{T}} } \right)
\left(
\frac{ 1+ e^{-\frac{p_0-\omega_-}{T}} }{ 1+ e^{-\frac{p_0-\omega_+}{T}}
} \right)
  \right] \right\} \theta(p_0^2 - {\bf p}^2 - (m_{X_1} + m_{X_2})^2) 
\nonumber \\ &+&
 g_X^2
\frac{-p_0^2 + {\bf p}^2}{8 \pi \omega_{X}({\bf p})}  
\left[1 - \frac{(m_{X_1} + m_{X_2})^2}{p_0^2 - {\bf p}^2}\right]
\frac{T}{|{\bf p}|} 
\nonumber \\
&\times & 
{\rm ln} \left[ \left(
  \frac{ 1+ e^{-\frac{\omega_+}{T}} }{ 1+ e^{-\frac{\omega_-}{T}} } \right)
\left(
\frac{ 1+ e^{-\frac{p_0+\omega_-}{T}} }{ 1+ e^{-\frac{p_0+\omega_+}{T}}
} \right)
  \right]  \theta(-p_0^2 + {\bf p}^2 + (m_{X_1} - m_{X_2})^2) 
\;.
\label{Gamma2}
\end{eqnarray}

\subsection{Decay of a fermion to a scalar boson and a fermion}

The last case is the decay of a fermion $X$  to another fermion $X_1$
and a scalar boson $X_2$. The Yukawa coupling among the fields is again
$g_X$. 
The imaginary part of the fermion field $X$
self-energy is in this case given by

\begin{equation}
{\rm Im} \Sigma_{X}(p_0, {\bf p}) = g_X^2   \left(1+e^{\beta p_0}
\right) \int \frac{d^4 k}{(2 \pi)^4} 
n_F(p_0-k_0) \rho_{X_1}(p_0-k_0, {\bf p}-{\bf k})  
n_B(k_0) \rho_{X_2}(k_0, {\bf k})\;.
\label{ImSigpsi}
\end{equation}

\noindent
Using the expressions for the free spectral functions, Eqs. (\ref{specsigma})
and (\ref{specpsi}), considering $p_0>0$, the branch cuts are
still given by~\cite{weldon} $-|m_{X_1}^2-m_{X_2}^2| \leq p_0^2 - {\bf p}^2 \leq
(m_{X_1}-m_{X_2})^2$ and $(m_{X_1}+m_{X_2})^2 \leq  p_0^2 -
{\bf p}^2 \leq \infty$. We obtain ${\rm Im} \Sigma_{X}= {\rm Im}
\Sigma_{X}^{\rm decay} + {\rm Im} \Sigma_{X}^{\rm LD}$, where
the decay term is

\begin{eqnarray}
{\rm Im} \Sigma_{\psi_\chi}^{\rm decay}(p_0, {\bf p}) &=& 2 \pi g_X^2 \int
\frac{d^3 k}{(2 \pi)^3} \frac{1}{4 \omega_{X_1}({\bf p}-{\bf k})\,
\omega_{X_2}({\bf k})
  } \left\{ \gamma^0 [p_0 -
  \omega_{X_2}({\bf k})] - {\bf \gamma}.{\bf p} +{\bf \gamma}.{\bf k} +
m_{X_1}\right\} \nonumber \\ &\times & \left[1+n_B(\omega_{X_2}({\bf
    k})) - n_F(p_0 -  \omega_{X_1}({\bf k})) \right] \delta(p_0 -
\omega_{X_1}({\bf p}-{\bf k})-
\omega_{X_2}({\bf k})  )\;,
\label{ImSigpsidecay}
\end{eqnarray}

\noindent
and the Landau damping term is

\begin{eqnarray}
{\rm Im} \Sigma_{\psi_\chi}^{\rm LD}(p_0, {\bf p}) &=& 2 \pi g_X^2 \int
\frac{d^3 k}{(2 \pi)^3} \frac{1}{4  \omega_{X_1}({\bf p}-{\bf k})
\omega_{X_2}({\bf k})
 } \left\{ -\gamma^0 [p_0 +
  \omega_{X_2}({\bf k})] + {\bf \gamma}.{\bf p} -{\bf \gamma}.{\bf k} -
m_{X_1}\right\} \nonumber \\ &\times & \left[n_F(p_0 +
  \omega_{X_1}({\bf k}))
+ n_B(\omega_{X_2}({\bf  k}))  \right] 
\delta(p_0 -\omega_{X_1}({\bf p}-{\bf k})+ \omega_{X_2}({\bf k})  )\;.
\label{ImSigpsiLD}
\end{eqnarray}

\noindent
The momentum integrals in the above expressions again follow  analogously to
the scalar boson case. This fermion self-energy can now be
expressed in the form
Eq. (\ref{ImSigmapsi}), from which explicit expressions for
$\tilde{\alpha}_0$, $\tilde{\alpha}_p$, and $\tilde{\alpha}_m$
can be identified.  These will be listed below.

The decay width in this case can be determined from the poles of the  dressed
fermion propagator,  $(\not\!p + m_{R, X} + i {\rm Im}
\Sigma_{X})^{-1}$, given by

\begin{equation}
\pm \omega_{X}({\bf p}) \pm i \left[ \omega_{X}({\bf p})
  \left(\tilde{\alpha}_p + \tilde{\alpha}_0\right)+
  \frac{m_{R,X}^2}{\omega_{X}({\bf p})} \left(\tilde{\alpha}_m
  - \tilde{\alpha}_p\right)\right] = \pm \omega_{X}({\bf p}) \pm i
\Gamma_{X}(p_0,{\bf p})\,,
\label{poles}
\end{equation}

\noindent
where the decay width of fermion $X$ to fermion $X_1$ and boson $X_2$
is found to be 

\begin{equation}
\Gamma_{X}(p_0,{\bf p})= \frac{1}{\omega_{X}({\bf p})}  \left[
  {\bf p}^2 \left( \tilde{\alpha}_p + \tilde{\alpha}_0\right)+
  m_{R,X}^2 \left(\tilde{\alpha}_m + \tilde{\alpha}_0
  \right) \right]\;.
\label{Gamma3}
\end{equation}
The explicit expressions for the $\tilde{\alpha}$ coefficients are

\begin{eqnarray}
p_0 \tilde{\alpha}_0 &=& \frac{g_X^2}{8 \pi |{\bf p}|}   \left[p_0 \left(
  \omega_+ - \omega_-\right) -\frac{1}{2}  \left(
  \omega_+^2 - \omega_-^2 \right) \right] \theta[p_0^2 - {\bf
    p}^2 - (m_{X_1} + m_{X_2})^2] \nonumber \\  
&+&  \frac{g_X^2 T}{8
  \pi |{\bf p}|} \left\{ \left(p_0 - \omega_+\right) {\rm ln} \left[
  \frac{1-e^{-\omega_+/T}}{1+e^{-|p_0-\omega_+|/T}}\right] -
\left(p_0 - \omega_-\right) {\rm ln} \left[
  \frac{1-e^{-\omega_-/T}}{1+e^{-|p_0-\omega_-|/T}} \right]
\right\} \theta[p_0^2 - {\bf p}^2 - (m_{X_1} + m_{X_2})^2]
\nonumber \\  
&+& \frac{g_X^2 T^2}{8 \pi |{\bf p}|}  \left\{  {\rm Li}_2  \left(
e^{-\omega_+/T} \right) -{\rm Li}_2  \left( e^{-\omega_-/T}
\right) +{\rm Li}_2  \left[- e^{-|p_0-\omega_+|/T} \right]
\right. \nonumber \\ 
&-& \left. {\rm Li}_2  \left[-
  e^{-|p_0-\omega_-|/T} \right] \right\} \theta[p_0^2 - {\bf p}^2 -
  (m_{X_1} + m_{X_2})^2] \nonumber \\ 
&+& \frac{g_X^2 T}{8 \pi |{\bf
    p}|} \left\{ \left(p_0 - \omega_+\right) {\rm ln} \left[
  \frac{1-e^{-\omega_+/T}}{1+e^{-(p_0+\omega_+)/T}}\right] -
\left(p_0 - \omega_-\right) {\rm ln} \left[
  \frac{1-e^{-\omega_-/T}}{1+e^{-(p_0+\omega_-)/T}} \right]
\right\} \theta[-p_0^2 + {\bf p}^2 + (m_{X_1} - m_{X_2})^2]
\nonumber \\  
&+& \frac{g_X^2 T^2}{8 \pi |{\bf p}|}  \left\{  {\rm Li}_2  \left(
e^{-\omega_+/T} \right) -{\rm Li}_2  \left( e^{-\omega_-/T}
\right) +{\rm Li}_2  \left[- e^{-(p_0+\omega_+)/T} \right]
\right. \nonumber \\ 
&-& \left. {\rm Li}_2  \left[-
  e^{-(p_0+\omega_-)/T} \right] \right\} \theta[-p_0^2 + {\bf p}^2 +
  (m_{X_1} - m_{X_2})^2] \;,
\label{alpha0}
\\  \tilde{\alpha}_p &=&\frac{g_X^2}{8 \pi |{\bf p}|}
\left[-\frac{p_0^2+{\bf p}^2- m_{X_1}^2+m_{X_1}^2 }{2 {\bf p}^2}
  \left(\omega_+ - \omega_-\right) +\frac{p_0}{2{\bf p}^2}
  \left( \omega_+^2 - \omega_-^2 \right) \right] \theta[p_0^2
  - {\bf p}^2 - (m_{X_1} + m_{X_2})^2] \nonumber \\  
&+&  \frac{g_X^2
  T }{8 \pi |{\bf p}|} \left\{  - \frac{\left(p_0^2+{\bf p}^2 -
  m_{X_1}^2+m_{X_2}^2  - 2 p_0 \omega_+\right)}{2 {\bf p}^2} {\rm ln}
\left[ \frac{1-e^{-\omega_+/T}}{1+e^{-|p_0-\omega_+|/T}}
  \right] \right.  \nonumber\\  
&+& \left. \frac{\left(p_0^2+{\bf
    p}^2- m_{X_1}^2+m_{X_2}^2  -  2 p_0 \omega_-\right)}{2
  {\bf p}^2} {\rm ln} \left[
  \frac{1-e^{-\omega_-/T}}{1+e^{-|p_0-\omega_-|/T}}\right]
\right\} \theta[p_0^2 - {\bf p}^2 - (m_{X_1} + m_{X_2})^2] 
\nonumber \\  
&+& \frac{g_X^2  T^2 p_0}{8 \pi |{\bf p}|^3} \left\{ {\rm Li}_2   \left(
e^{-\omega_+/T} \right) - {\rm Li}_2   \left( e^{-\omega_-/T}
\right) + {\rm Li}_2  \left[- e^{-|p_0-\omega_+|/T} \right]
\right. \nonumber \\ 
&-& \left. {\rm Li}_2  \left[-
  e^{-|p_0-\omega_-|/T} \right] \right\} \theta[p_0^2 - {\bf p}^2 -
  (m_{X_1} + m_{X_2})^2] \nonumber \\  
&+&  \frac{g_X^2 T }{8 \pi
  |{\bf p}|} \left\{  - \frac{\left(p_0^2+{\bf p}^2-
  m_{X_1}^2+m_{X_2}^2  - 2 p_0 \omega_+\right)}{2 {\bf p}^2} {\rm ln}
\left[ \frac{1-e^{-\omega_+/T}}{1+e^{-(p_0+\omega_+)/T}}
  \right] \right.  \nonumber\\  
&+& \left. \frac{\left(p_0^2+{\bf
    p}^2- m_{X_1}^2 +m_{X_2}^2  -  2 p_0 \omega_-\right)}{2
  {\bf p}^2} {\rm ln} \left[
  \frac{1-e^{-\omega_-/T}}{1+e^{-(p_0+\omega_-)/T}}\right]
\right\} \theta[-p_0^2 + {\bf p}^2 + (m_{X_1} - m_{X_2})^2] 
\nonumber \\  
&+& \frac{g_X^2  T^2 p_0}{8 \pi |{\bf p}|^3} \left\{ {\rm Li}_2   \left(
e^{-\omega_+/T} \right) - {\rm Li}_2   \left( e^{-\omega_-/T}
\right) + {\rm Li}_2  \left[- e^{-(p_0+\omega_+)/T} \right]
\right. \nonumber \\ 
&-& \left. {\rm Li}_2  \left[-
  e^{-(p_0+\omega_-)/T} \right] \right\} \theta[-p_0^2 + {\bf p}^2 +
  (m_{X_1} - m_{X_2})^2]\;,
\label{alphap}
\\  \tilde{\alpha}_m &=&\frac{g_X^2}{8 \pi |{\bf
    p}|}\frac{m_{X_1}}{m_{R,X}}   \left( \omega_+ -
\omega_-\right) \theta[p_0^2 - {\bf p}^2 - (m_{X_1} +
  m_{X_2})^2] \nonumber \\  
&+&  \frac{g_X^2}{8 \pi |{\bf p}|} T
\frac{m_{X_1}}{m_{R,X}} \left\{  {\rm ln} \left[
  \frac{1-e^{-\omega_+/T}}{1+e^{-|p_0-\omega_+|/T}}\right]
-{\rm ln} \left[
  \frac{1-e^{-\omega_-/T}}{1+e^{-|p_0-\omega_-|/T}}\right]
\right\} \theta[p_0^2 - {\bf p}^2 - (m_{X_1} + m_{X_2})^2]  
\nonumber \\ 
&+&  \frac{g_X^2}{8 \pi |{\bf p}|} T
\frac{m_{X_1}}{m_{R,X}} \left\{  {\rm ln} \left[
  \frac{1-e^{-\omega_+/T}}{1+e^{-(p_0+\omega_+)/T}}\right]
-{\rm ln} \left[
  \frac{1-e^{-\omega_-/T}}{1+e^{-(p_0+\omega_-)/T}}\right]
\right\} \theta[-p_0^2 + {\bf p}^2 + (m_{X_1} - m_{X_2})^2]\;,
\label{alpham}
\end{eqnarray}
where ${\rm Li}_2(x)$ is the dilogarithm function,
\begin{equation}
{\rm Li}_2(x) = \sum_{n=1}^{\infty} \frac{x^n}{n^2}\;,
\end{equation}
and $\omega_\pm$ are given again by Eq. (\ref{omegapm}).

\acknowledgements

A.B. acknowledges support from the STFC.  M.B. is partially supported by the
M.E.C. under contract FIS2007-63364 and by the Junta de Andaluc\'{\i}a group
FQM 101.   R.O.R. would like to thank the hospitality of the School of 
Physics and Astronomy at the
University of Edinburgh which during his visit this work has started. R.O.R 
is partially supported by Conselho Nacional de
Desenvolvimento Cient\'{\i}fico e Tecnol\'ogico (CNPq - Brazil) and by SUPA
during the realization of this~work~in~the~UK.

\end{document}